\documentclass[twocolumn]{aastex631}
\usepackage{geometry}
\usepackage{amsmath}
\usepackage{graphicx}
\usepackage{color}
\usepackage{dcolumn}

\usepackage{tikz}
\usetikzlibrary{matrix}
\usepackage{natbib}
\definecolor{brickred}{rgb}{0.8, 0.25, 0.33}
\newcommand{\quotes}[1]{``#1''}
\begin{document}

\title{The observed phase space of mass-loss history from massive stars based on radio observations of a large supernova sample}

\author[0000-0003-0466-3779]{Itai Sfaradi}
\affiliation{Racah Institute of Physics, The Hebrew University of Jerusalem, Jerusalem 91904, Israel}
\author[0000-0002-5936-1156]{Assaf Horesh}
\affiliation{Racah Institute of Physics, The Hebrew University of Jerusalem, Jerusalem 91904, Israel}
\author{Rob Fender}
\affiliation{Astrophysics, Department of Physics, University of Oxford, Keble Road, Oxford, OX1 3RH, UK}
\author{Lauren Rhodes}
\affiliation{Astrophysics, Department of Physics, University of Oxford, Keble Road, Oxford, OX1 3RH, UK}
\author{Joe Bright}
\affiliation{Astrophysics, Department of Physics, University of Oxford, Keble Road, Oxford, OX1 3RH, UK}
\author{David Williams-Baldwin}
\affiliation{Jodrell Bank Centre for Astrophysics, School of Physics and Astronomy, The University of Manchester, Manchester, M13 9PL, UK}
\author{Dave A. Green}
\affiliation{Astrophysics Group, Cavendish Laboratory, 19 J. J. Thomson Ave., Cambridge CB3 0HE, UK}

\begin{abstract}
In this work we study the circumstellar material (CSM) around massive stars, and the mass-loss rates depositing this CSM, using a large sample of radio observations of 325 core-collapse supernovae (CCSNe; only $\sim 22 \%$ of them being detected). This sample comprises both archival data and our new observations of 99 CCSNe conducted with the AMI-LA radio array in a systematic approach devised to constrain the mass-loss at different stages of stellar evolution. In the SN-CSM interaction model, observing the peak of the radio emission of a SN provides the CSM density at a given radius (and therefore mass-loss rate that deposited this CSM). On the other hand, limits on the radio emission, and/or on the peak of the radio emission provide a region in the CSM phase space that can be ruled out. Our analysis shows discrepancy between the values of mass-loss rates derived from radio-detected and radio-non-detected SNe. Furthermore, we rule out mass-loss rates in the range of $2 \times 10^{-6} - 10^{-4} \, \rm M_{\odot} \, yr^{-1}$ for different epochs during the last 1000 years before the explosion (assuming wind velocity of $10 \, \rm km \, s^{-1}$) for the progenitors of $\sim 80\%$ of the type II SNe in our sample. In addition, we rule out the ranges of mass-loss rates suggested for red supergiants for $\sim 50 \%$ of the progenitors of type II SNe in our sample. We emphasize here that these results take a step forward in constraining mass-loss in winds from a statistical point of view.
\end{abstract}

\smallskip
\section{Introduction}
\label{sec:intro}

Mass loss from massive stars, stars with $M_{\rm Zero \, Age \, Main \, Sequence} \geq 8 \: \rm{M_{\odot}}$, can be significant and affect their luminosity, lifetime, composition, and mass. Towards the end of its life, mass loss from a massive star can strongly influence the fate of the star and the resulting SN \citep{Annu_rev_Smith_2014}. Various models and observations suggest ranges of mass-loss rates for massive stars during different phases of their evolution (e.g., \citealt{de_jager_1988, vink_2001, Annu_rev_Langer_2012, Annu_rev_Smith_2014, vink_2022, irani_2023}). However, mass loss at the last stages of stellar evolution is still unconstrained empirically.

Type II (hydrogen-rich) supernovae (SNe) and their variations (II-L, II-P, IIn; excluding type IIb), and stripped-envelope SNe (type IIb, Ib, and Ic; hydrogen-poor) are associated with the death of massive stars as they reach the evolutionary point when nuclear burning can no longer provide support against the star's own gravity. This stellar death can result in a core-collapse SN (CCSN) explosion. In these explosions, stellar material is ejected at very high velocities reaching $\sim 10,000 \, \rm km \, s^{-1}$ for type II SNe, and even higher ejecta velocities for stripped-envelope SNe \citep{Filippenko_1997, SNe_handbook, Chevalier_2006}. The interaction between the SN ejecta and the circumstellar material (CSM) generates the radio emission \citep{Chevalier_1981,Chevalier_1998,Weiler_2002}. The CSM closest to the star has been deposited via mass-loss processes (e.g., stellar winds and eruptive mass ejection), just before the explosion. Thus, early radio observations of young SNe reveal the environment of the progenitor at its final evolutionary stages, and its mass-loss rate just before the explosion. 

In the last 4 decades, an increasing number of SNe have been detected in radio, revealing surprising variations in spectra and light curves. While most stripped-envelope SNe exhibit clear self-absorbed Synchrotron spectrum \citep{Chevalier_1998}, most type II SNe point towards a spectrum associated with free-free absorption (e.g., SN\,1979C \citealt{SN1979C}, SN\,2023ixf; \citealt{SN2023ixf}). The radio luminosity-rise time relation of CCSNe was studied by \cite{bietenholz_2020} on a large sample of radio observation of CCSNe. Their analysis of all the SNe in their sample (detected, and non-detected in radio wavelengths) suggests that the peak luminosity of type Ib, Ic, and II SNe is $\sim 2 \times 10^{25} \, \rm erg \, s^{-1} \, Hz^{-1}$, and that type Ib, and Ic SNe rise-time is $12-15$ days after the SN explosion, shorter than the $\sim 30 - 100$ days which was inferred for type II SN. Since type IIn SNe present narrow lines in their optical spectrum, associated with interaction with a dense CSM, they were excluded from this analysis. In addition, type IIb SNe, presenting a transitional optical spectrum (initially showing hydrogen lines in the optical spectrum that disappear at late times), were also excluded from the analysis above. Based on these results, \cite{bietenholz_2020} inferred a mean mass-loss rate for the progenitors of type Ib and Ic SNe of $\sim 4 \times 10^{-6} \, \rm M_{\odot} \, yr^{-1}$ (for an assumed constant mass-loss rate in steady winds of $1000 \, \rm km \, s^{-1}$). For the progenitors of type II SNe (excluding type IIn), the inferred mean of mass-loss rate is $\sim 10^{-7}\, \rm M_{\odot} \, yr^{-1}$ (for winds of $10 \, \rm km \, s^{-1}$).

Some SNe, for example SN\,2014C \citep{SN2014C_2}, exhibit multiple peaks in their radio light curve suggesting that the SN ejecta interacts with two separate CSM shells deposited in two separate mass-loss stages. Some SNe (e.g., SN\,2003L; \citealt{SN2003L}, SN\,2019oys; \citealt{SN2019oys}) exhibit broad radio spectral peaks which are associated with CSM inhomogeneities caused by variations in the distribution of relativistic electrons and/or the magnetic field strength within the synchrotron source. Fast shockwaves ($0.2-0.3 \, c$) traveling in a dense environment have been captured when observations were taken early enough (e.g. the type Ic SNe - SN2002ap \citealt{SN2002ap}; PTF12gzk \citealt{PTF12gzk}) and can link normal Type Ic to relativistic ones \citep{Margutti_2014}.

A typical assumption when studying the radio emission from CCSNe is that there is equipartition between the energy deposited in the relativistic electrons and in the magnetic fields \citep{scott_readhead_1977, Chevalier_1998, barniol_duran_2013}. Combined radio and X-ray observations (e.g. SN\,2011dh; \citealt{SN2011dh_1}; \citealt{SN2011dh_3}; \citealt{SN2011dh_2}, SN\,2013df; \citealt{SN2013df}, SN\,2020oi; \citealt{SN2020oi_horesh}) displayed deviation from this equipartition (with a larger fraction of energy deposited in the relativistic electrons). Other early multi-wavelength observations have shown the importance of electron cooling, including SN\,2012aw \citep{SN2012aw}, in which the steep radio spectrum observed early on showed a significant inverse Compton cooling at frequencies above $1$ GHz. In addition to this large variety of detected SNe, there are a lot more SNe that were observed and were not detected at radio wavelengths. A non-detection of an SN in radio can rule out a range of mass-loss rates from the progenitor star. 

In this paper, we probe the CSM around massive stars (and the mass-loss rates depositing this CSM) from a population point of view, using radio observations of both radio-detected and non-detected CCSNe. We start by introducing the SN-CSM interaction model in \S\ref{sec:Modeling}. In \S\ref{sec:sample} we describe the large sample of CCSNe observed in radio wavelengths, with new observations of CCSNe first introduced here. We analyze this sample in \S\ref{sec: detections} and \S\ref{sec: non-detections} according to the model described in \S\ref{sec:Modeling}. In \S\ref{sec: caveats} we test our model assumptions and their impact on our conclusions. $\S$\ref{sec:summary} is for conclusions.

\section{SN-CSM interaction model}
\label{sec:Modeling}

In the SN-CSM interaction model (\citealt{Chevalier_1981}) the SN ejecta drives a shockwave into the CSM. At the shock front, particles are accelerated to relativistic velocities and gyrate in the presence of a magnetic field. These relativistic particles give rise to synchrotron emission, which is usually brightest at radio frequencies. While \cite{Chevalier_1998} showed that the early radio emission is fairly well described by a synchrotron self-absorption (SSA) model, free--free absorption (FFA; \citealt{Weiler_2002}) can also take part. Next, we describe the model for the SSA spectrum and connect it to the density of the CSM and mass-loss rate from the progenitor. Then, we show the full model we use by introducing external FFA to the SSA spectrum.

In this model, the relativistic particles have an energy density distribution of $N(E) = N_{\rm 0} E^{-p}$. As described in Eq. (1) in \cite{Chevalier_1998}, the flux density from a SSA spectrum at a frequency $\nu$ is given by
\begin{align}
\label{eq: full_ssa}
    \nonumber & F_{\rm \nu} = \frac{\pi R^2}{D^2} \left( \frac{\nu}{\nu_{\rm 1}}\right)^{5/2} \times \\ 
    &\left( 1- \exp\left[ -\left( \frac{\nu}{\nu_{\rm 1}}\right)^{-(p+4)/2} \right] \right) \\
    \nonumber & \rm{where} \\
    \nonumber & \nu_{\rm 1} = 2c_{\rm 1} \left( \frac{4}{3} f R c_{\rm 6} N_{\rm 0} \right)^{2/(p+4)} B^{(p+2)/(p+4)},
\end{align} 
$R$ is the radius of the radio-emitting shell, $B$ is the magnetic field strength, $f$ is the emission filling factor and $D$ is the distance to the SN. The constants $c_1$, $c_5$, and $c_6$ can be found in \citep{pacholczyk_1970}. Since the total energy density in relativistic electrons is $\int^{\infty} _{E_{\rm l}} N(E) \, {\rm d}E$, where $E_{\rm l}$ is the electron rest mass energy\footnote{Here we assumed that electron cooling effects are not important at the observed frequency. This is not always the case, see e.g. SN\,2020oi; \cite{SN2020oi_horesh}, SN\,2019oys; \cite{SN2019oys}, SN2012aw; \cite{SN2012aw}.}, the equipartition assumption \citep{scott_readhead_1977,Chevalier_1998} gives 
\begin{align}
    N_0 = \frac{f_{\rm eB}B^2 (p-2)E_l ^{p-2}}{8 \pi}.
\end{align}

\cite{Chevalier_2006} showed that one can estimate the radius of the emitting shell and the magnetic field strength at that time given the radio spectral peak flux density, $F_{\nu_a}$, and frequency, $\nu_a$. The radius is given by
\begin{align}
    R = & \left[ \frac{6 c_6 ^{p+5} F_{\nu_a} ^{p+6} D^{2p+12}}{f_{\rm eB} f \left( p-2 \right) E_{\rm l} ^{p-2} \pi^{p+5} c_5 ^{p+5}} \right]^{\frac{1}{2p+13}} \left( \frac{\nu_{\rm a}}{2 c_1} \right)^{-1},
    \label{eq: Radius_chevalier}
\end{align}
and the magnetic field strength is
\begin{align}
    B = & \left[ \frac{36 \pi^3 c_5}{f_{\rm eB} ^2 f^2 \left(p-2\right)^2 E_{\rm l} ^{2(p-2)} c_6 ^3 F_{\nu_a} D^2} \right]^{\frac{2}{2p+13}} \left( \frac{\nu_{\rm a}}{2 c_1} \right),
    \label{eq: Magnetic_chevalier}
\end{align}
where $f_{\rm eB}$ is the ratio between the fraction of shock wave energy deposited in the relativistic electrons ($\epsilon_{\rm e}$) to the fraction of shock wave energy converted to the magnetic field ($\epsilon_{\rm B}$).

The temporal evolution of the radio spectrum is determined by the evolution of the radius and magnetic field with time. A typical assumption is of a free expansion of the shock which dictates a constant shock velocity, $v_{\rm sh} = R/t$, where $t$ is the time since the explosion. Assuming that a fraction $\left( \epsilon_{\rm B} \right)$ of the post-shock thermal energy density, $U_{\rm ps} = \frac{9}{8} \rho_{\rm CSM} v_{\rm sh}^2$ (the factor of $9/8$ in the post-shock thermal energy arises from the Rankine–Hugoniot jump conditions, see discussion in the appendix of \citealt{SN2004cc_demarchi} for more details) where $\rho_{\rm CSM}$ is the density of the CSM, is transferred to the magnetic field energy density, $U_{\rm B} = B^2/8 \pi$, gives 
\begin{align}
\label{eq: b_rho_relation}
    B = \sqrt{9 \pi \epsilon_{\rm B} \rho_{\rm CSM} v_{\rm sh}^2}.    
\end{align}
Therefore, the radial shape of the density profile and the evolution of the shock with time determine the temporal evolution of the magnetic field strength. A typical assumption for the formation of the CSM is via a constant mass-loss rate in steady winds (e.g. stellar winds). This scenario leads to a density profile of
\begin{align}
\label{eq: rho_density_param}
    \rho_{\rm CSM} \left( r \right) = \frac{\dot{M}/v_{\rm w}}{4 \pi r^2},    
\end{align}
where $\dot{M}$ is the mass-loss rate and $v_{\rm w}$ is the wind velocity.

External FFA is important when the external optical depth, $\tau_{\rm ff}$, is of the order of unity (as seen in numerous cases, e.g., SN\,1979C; \citealt{SN1979C}, SN\,2013df; \citealt{SN2013df}, SN\,2019oys; \citealt{SN2019oys}, SN\,2023ixf; \citealt{SN2023ixf}). This will attenuate the flux density shown in Eq. \ref{eq: full_ssa} by a factor of $e^{-\tau_{\rm ff}}$ where the optical depth of FFA is (see e.g., \citealt{SN2011dh_2})
\begin{align}
    \nonumber \tau_{\rm ff} = & 0.76 \left( \frac{\dot{M} \left[ 10^{-6}\,{\rm M_{\odot} \, yr^{-1}} \right] }{v_{\rm w} \left[ 10\, \rm{km \, s^{-1}} \right]} \right)^{2} \left( \frac{T_{\rm e}}{10^5\,\rm{K}} \right)^{-1.35} 
    \\
    &\left( \frac{v_{\rm sh}}{10^4\,\rm{km \, s^{-1}}} \right)^{-3} \left( \frac{t}{10\,\rm{days}} \right)^{-3} \left( \frac{\nu}{5\,\rm{GHz}} \right)^{-2.1},
    \label{eq: FFA optical depth}
\end{align}
where $T_{\rm e}$ is the temperature of the electrons in the wind.

In the following sections, we analyze the radio data shown in $\S$\ref{sec:sample} in light of the SN-CSM interaction model described above under several assumptions. First, throughout the entire analysis, we assumed that $p=3$ for stripped envelope SNe, and $p=2.4$ for Type II SNe. This assumption is based on radio observations of the optically thin regime for various CCSNe (see Table 1 in \citealt{Chevalier_1998}, Table 2 in \citealt{Weiler_2002}, and the discussion in \citealt{bietenholz_2020}). We discuss the implications of changing this assumption in \S\ref{subsec:powerlaw}). Another assumption is that of equipartition, and specifically $\epsilon_{\rm e} = \epsilon_{\rm B} = 0.1$. While there is great theoretical and observational motivation for the equipartition assumption (e.g., \citealt{scott_readhead_1977, Chevalier_1998, barniol_duran_2013}) the exact values of $\epsilon_{\rm e}$ and $\epsilon_{\rm B}$ are not well understood observationally (we discuss the implications of this assumption in  \S\ref{subsec:equipartition}. We also assumed that the emission filling factor is $f=0.5$. This assumption means that the emitting volume is $\sim 20 \%$ of the volume engulfed by the shock front. Finally, for the temperature of the CSM we assumed $T_{\rm e} = 10^5$ K but note that the true value for the temperature may vary (see e.g. \citealt{lundqvist_1988,chevalier_fransson_2017}).

\section{Radio Supernovae sample}
\label{sec:sample}

Over the last four decades, astronomers conducted and reported radio observations of CCSNe in different wavelengths (typically in the GHz band) and different time scales after the stellar explosion. We assembled a sample of radio observations of CCSNe using data from literature, online databases, and reports (e.g. The Open SNe Catalogue\footnote{https://sne.space/}; ATels\footnote{http://www.astronomerstelegram.org/}; GCN Circulars\footnote{https://gcn.gsfc.nasa.gov/gcn3\_archive.html}). In addition to these previously reported observations, we present here for the first time radio observations of 99 CCSNe conducted with the Arcminute Micro-Kelvin Imager - Large Array (AMI-LA; \citealt{zwart_2008,hickish_2018}) as part of our systematic observing campaign. 

\subsection{The Arcminute Micro-Kelvin Imager - Large Array}

AMI-LA is a radio interferometer comprised of eight, 12.8-m diameter, antennas producing 28 baselines that extend from 18-m up to 110-m in length and operate around a central frequency of 15.5 GHz with a 5 GHz bandwidth, divided into eight channels. For each observation, initial data reduction, flagging, and calibration of the phase and flux, were carried out using \texttt{reduce\_dc}, a customized AMI-LA data reduction software package \citep{perrott_2013}. Phase calibration was conducted using short interleaved observations of the phase calibrator, while daily observations of 3C286 were used for absolute flux calibration. Additional flagging was performed using CASA \citep{mcmullin_casa}. 

In our observations with the AMI-LA we took a systematic approach. We conducted the first observations as soon as possible (given technical limitations) following a classification of a CCSNe. Then, we aimed to monitor each SN (even in the case of a radio non-detection) on time scales of: a week, a month, three months, half a year, a year, and a year and a half after the explosion. This was done to probe the CSM at different regions around the CCSNe, and in the case of consecutive upper limits, constraining the possible mass-loss rates from the progenitor massive star at different epochs of stellar evolution.

\subsection{The Complete Sample}
The resulting SNe sample is comprised of 325 SNe, 177 of which are Type II SNe (including its various sub-types, e.g.\ II-P, II-L, and IIn; Pec is for Peculiar SNe) and 148 are stripped-envelope SNe (Types Ib, Ic, and IIb; BL is for Broad-Line SNe). Within this set, 72 have radio detections, while 253 have only upper limits on the radio flux density. From this search for radio emission from CCSNe we find out that $\sim 78 \%$ of CCSNe in the near universe ($90 \%$ of the CCSNe in our sample are at a distance of up to $100$ Mpc) were not detected in radio wavelengths. See Table \ref{table:Database summary - 1} for the SNe in our sample, and Table \ref{table:Limits_summary} in the appendix for the radio upper limits in our sample. Finally, Table \ref{table:Peaks_summary} in the appendix for the peak flux density of radio detected SNe (as described in detail in \S\ref{sec: detections}).

\begin{deluxetable*}{cccccp{3cm}}
\tablecaption{Summary of the properties of CCSNe in our sample \label{table:Database summary - 1}}
\tablehead{
\colhead{Name} &
\colhead{Type} &
\colhead{Explosion Date} &
\colhead{Distance} &
\colhead{Detected} &
\colhead{Reference(s)} \\
\colhead{} &
\colhead{} &
\colhead{[DD/MM/YYYY]} &
\colhead{[Mpc]} &
\colhead{[Y/N]} &
\colhead{} 
}
\startdata
SN1979C & IIL & 06/04/1979 & 16.2 & Y & \scriptsize{\parbox[c]{3cm}{\cite{SN1979C} \\ \cite{bietenholz_2020}}} \\ [0.5ex]
SN1980K & II & 14/10/1980 & 5.5 & Y & \scriptsize{\parbox[c]{3cm}{\cite{SN1980K} \\ \cite{bietenholz_2020}}} \\ [0.5ex]
SN1980O & II & 30/12/1980 & 17.9 & N & \scriptsize{\parbox[c]{3cm}{\cite{bietenholz_2020}}} \\ [0.5ex]
SN1981K & II & 31/07/1981 & 7.3 & Y & \scriptsize{\parbox[c]{3cm}{\cite{SN1981K} \\ \cite{bietenholz_2020}}} \\ [0.5ex]
SN1982F & IIP & 24/02/1982 & 6.2 & N & \scriptsize{\parbox[c]{3cm}{\cite{bietenholz_2020}}} \\ [0.5ex]
SN1986J & II & 14/03/1983 & 10.0 & Y & \scriptsize{\parbox[c]{3cm}{\cite{SN1986J} \\ \cite{bietenholz_2020}}} \\ [0.5ex]
SN1983N & Ib & 29/06/1983 & 4.9 & Y & \scriptsize{\parbox[c]{3cm}{\cite{SN1983N} \\ \cite{bietenholz_2020}}} \\ [0.5ex]
SN1984E$^a$ & IIL & 26/03/1984 & 17.54 & N & \scriptsize{\parbox[c]{3cm}{\cite{SN1984E} \\ \cite{bietenholz_2020}}} \\ [0.5ex]
\enddata
\tablecomments{In the \quotes{Name} column, $^a$ indicates when the discovery date from the open SNe catalog is taken as an explosion date since no other data was available. The distance for these objects is inferred directly from redshift, assuming $H_0=70 \, \rm km\,s^{-1}/Mpc$ and $\Omega_m = 0.25$. Where $^a$ is not mentioned, the distance and explosion date were taken from the referenced article. All SNe first presented here are marked in the \quotes{Reference} column as \quotes{This work}. For these SNe the explosion date is based on the discovery date from their TNS page, and their distance is inferred directly from redshift using the same assumption mentioned above. $^b$ indicates when the radio peak is taken directly from the literature and not by the fitting process described in \S\ref{sec: detections}.\\
Table \ref{table:Database summary - 1} is published in its entirety in the machine-readable format. A portion is shown here for guidance regarding its form and content.
}
\tablerefs{
\citet{Dessart_2008}
\citet{PTF10vgv}; 
\citet{PTF11iqb}; 
\citet{PTF12gzk}; 
\citet{SN1979C}; 
\citet{SN1980K}; 
\citet{SN1981K}; 
\citet{SN1983N}; 
\citet{SN1984E}; 
\citet{SN1985L}; 
\citet{SN1986E}; 
\citet{SN1986J}; 
\citet{SN1988Z}; 
\citet{SN1990B}; 
\citet{SN1992ad}; 
\citet{SN1993J}; 
\citet{SN1993J_pooley}; 
\citet{SN1994I}; 
\citet{SN1995N}; 
\citet{SN1997eg_1}; 
\citet{SN1997eg_2}; 
\citet{SN1998S_1}; 
\citet{SN1998S_2}; 
\citet{SN1998bw}; 
\citet{SN1999em_1}; 
\citet{SN1999em_2}; 
\citet{SN1999gi}; 
\citet{SN2000ft}; 
\citet{SN2001em_1}; 
\citet{SN2001em_2}; 
\citet{SN2001gd}; 
\citet{SN2001ig}; 
\citet{SN2002ap}; 
\citet{SN2003L}; 
\citet{SN2003bg}; 
\citet{SN2003ed}; 
\citet{SN2003jd}; 
\citet{SN2004am}; 
\citet{SN2004dj}; 
\citet{SN2004dk_SN2004gq_SN2004cc}; 
\citet{SN2004et_1}; 
\citet{SN2004et_2}; 
\citet{SN2004et_3}; 
\citet{SN2004gk_1}; 
\citet{SN2004gk_2}; 
\citet{SN2005V}; 
\citet{SN2005bf}; 
\citet{SN2005cs_2}; 
\citet{SN2005da}; 
\citet{SN2005ip}; 
\citet{SN2005kd}; 
\citet{SN2006bp_2}; 
\citet{SN2006gy_1}; 
\citet{SN2006gy_2}; 
\citet{SN2006gy_3}; 
\citet{SN2006gy_4}; 
\citet{SN2006gy_5}; 
\citet{SN2006gy_6}; 
\citet{SN2006jc}; 
\citet{SN2006jd}; 
\citet{SN2007Y}; 
\citet{SN2007gr}; 
\citet{SN2007pk}; 
\citet{SN2007rt}; 
\citet{SN2007uy_2}; 
\citet{SN2008D_SN2007uy_1}; 
\citet{SN2008ax_1}; 
\citet{SN2008ax_2}; 
\citet{SN2008bk_1}; 
\citet{SN2008bk_2}; 
\citet{SN2008bk_3}; 
\citet{SN2008bo}; 
\citet{SN2008gm}; 
\citet{SN2008ij_1_SN2009H}; 
\citet{SN2008ij_2}; 
\citet{SN2008in_1}; 
\citet{SN2008in_2}; 
\citet{SN2008in_3}; 
\citet{SN2008ip}; 
\citet{SN2008iz}; 
\citet{SN2009au}; 
\citet{SN2009bb}; 
\citet{SN2010O_SN2010P}; 
\citet{SN2010br}; 
\citet{SN2010cu}; 
\citet{SN2010jl}; 
\citet{SN2011dh_1}; 
\citet{SN2011dh_2}; 
\citet{SN2011ei}; 
\citet{SN2011hp}; 
\citet{SN2012ap}; 
\citet{SN2012au}; 
\citet{SN2012aw}; 
\citet{SN2013ai}; 
\citet{SN2013ej_1}; 
\citet{SN2013ej_2}; 
\citet{SN2013fs}; 
\citet{SN2013ge}; 
\citet{SN2014C_1}; 
\citet{SN2014C_2}; 
\citet{SN2014bc_2_SN2014bi}; 
\citet{SN2015G}; 
\citet{SN2018ec_1}; 
\citet{SN2018ec_2}; 
\citet{SN2018lab}; 
\citet{SN2019oys}; 
\citet{SN2020oi_horesh}; 
\citet{SN2020qmp}; 
\citet{Weiler_2002}; 
\citet{bietenholz_2020}; 
\citet{iPTF13bvn_1}; 
\citet{iPTF13bvn_2}; 
\citet{iPTF13bvn_3}; 
\citet{soderberg_2006_sample}
}
\end{deluxetable*}
\normalsize

In the following sections, we probe the phase space of density profiles and mass-loss rates from massive stars in light of the model presented in \S\ref{sec:Modeling} using the radio observations of CCSNe in our sample. We first analyze radio-detected SNe and discuss the results in terms of shock velocities and mass-loss rate in $\S$\ref{sec: detections}. Next, in $\S$\ref{sec: non-detections}, we use the entire sample of SNe (both radio-detected and radio-non-detected SNe) to probe the CSM density around CCSNe and constrain the phase space of the mass-loss rate from massive stars.

\section{Analysis of SNe with radio detections}
\label{sec: detections}

The SN-CSM model described in \S\ref{sec:Modeling} suggests that by observing the radio peak flux density we can infer the velocity of the shock traveling in the CSM, and the mass-loss rate divided by wind velocity. Thus, in the following section, we analyze SNe (from the sample shown in $\S$\ref{sec:sample}) which were detected in radio, by deriving their spectral peak flux density, frequency, and time since the explosion. To accomplish that, we fit the following two power-law spectra to the SN radio data (see Eq. $4$ in \citealt{Chevalier_1998})
\begin{align}
    \nonumber F_{\nu} \left( t \right) = & 1.582 \, F_{\rm p} \left( \frac{t}{t_a} \right)^a \left( \frac{\nu}{\nu_a} \right)^{5/2} \times \\
    &\left( 1 - \exp{^{- \left( \frac{t}{t_a} \right)^{-\left(a+b\right)} \left( \frac{\nu}{\nu_a} \right)^{-\left( p+4 \right)/2}}} \right)
    \label{eq: chev_eq_4}
\end{align}
where $\nu_{\rm a}$ is the frequency where the optical depth is around unity at a time $t_{\rm a}$ after the explosion, and $F_{\rm p}$ is the flux density at this time, and $a$ and $b$ are the time evolution power-law indices of the optically thick and thin regimes, respectively. $F_{\nu_{a}} = 1.582 \, F_{\rm p}$ is the flux density at the intersection between the optically thin and thick regime (used in Eq. \ref{eq: Radius_chevalier} and \ref{eq: Magnetic_chevalier}). For a shockwave with a constant expansion velocity traveling in a radial density profile of $r^{-2}$ we get $a=2.5$ and $b=1$ \citep{Chevalier_1998}. When the spectrum of a single SN, at some time $t$, exhibits both an optically thick and thin regime, we fitted the spectrum with Eq. \ref{eq: chev_eq_4} assuming $t=t_a$. $F_{\rm p}$, $\nu_a$ and $p$ were treated as free parameters.

For SNe with poor spectral coverage, and a detailed light curve in only one radio frequency, $\nu$, we assumed that a turnover in their radio light curve (if observed) is due to a spectral transition between the optically thick and thin regimes. We treat such SNe by fitting Eq. \ref{eq: chev_eq_4} to their radio light curve. The fit free parameters were $F_{\rm p}$, $t_a$, $a$ and $b$, and we assumed $\nu=\nu_a$. The results of these fits and the fits of the spectra (with a total of for $45$ SNe) are summarized in Table \ref{table:Peaks_summary}. In three special cases (SN\,2002ap, SN\,2008ax, and SN\,2012ap), as described in Table \ref{table:Database summary - 1}, we used the spectral peak reported in previous papers.

\subsection{Chevalier's diagram}
\label{subsec: chevalier_diagram}

\begin{figure*}
\begin{center}
\includegraphics[width=\textwidth]{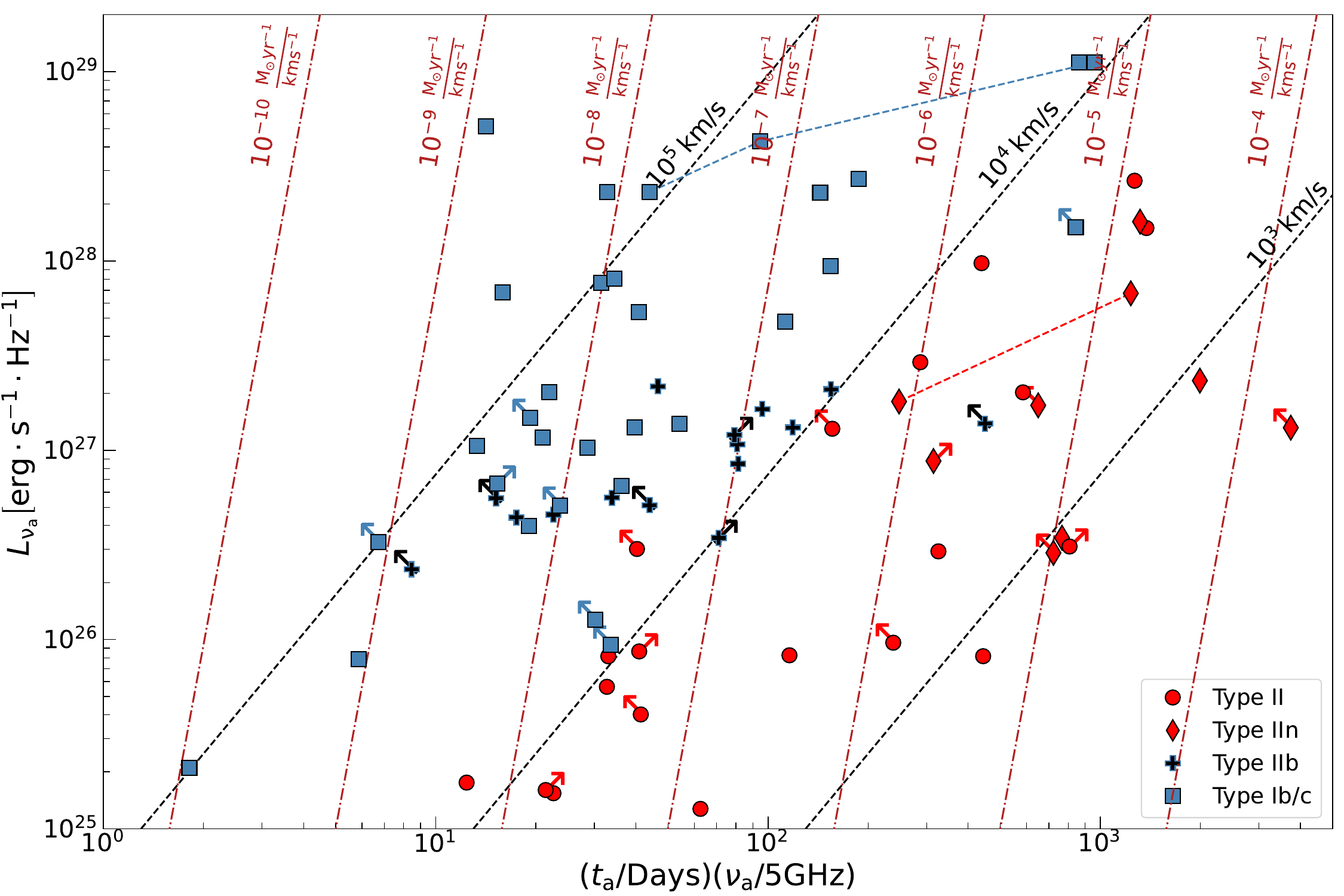}
\caption{Chevalier diagram showing the peak spectral radio luminosity as a function of $t_a \nu_a$. The peak of each SN was obtained by fitting Eq.4 in \cite{Chevalier_1998} to the spectrum at a specific time, or to the light curve of a specific frequency (see $\S$\ref{sec: detections}). The arrows are for an SN whose peak was not observed, and the direction of the arrow indicates whether the peak is in later or earlier times than the time of the highest observed flux density. Points connected with dashed lines represent multiple peaks in the radio light curves for a single SNe. Lines of equal shock velocity (dashed black lines), and of the CSM density parameter (dash-dot red lines) are plotted according to Eq. \ref{eq: Radius_chevalier}, \ref{eq: Magnetic_chevalier}, \ref{eq: b_rho_relation}, and \ref{eq: rho_density_param} under the assumption that the peak is due to SSA, and of $p=3$, $f=0.5$ and equipartition ($\epsilon_{\rm e} = \epsilon_{\rm B} = 0.1$). Although we derive the mass-loss rate for type II SNe assuming $p=2.4$, for the sake of comparison only, and specifically for this Chevalier's diagram, we present the lines of equal shock velocities and mass-loss rates assuming $p=3$.}
\label{fig: Chevalier_diagram}
\end{center}
\end{figure*}

The phase space of the peak radio luminosity, $L_{\rm \nu_{\rm a}} = 4 \pi D^2 F_{\rm \nu_{\rm a}}$, and the time of the observation times the observed peak frequency, $t_{\rm a}\nu_{\rm a}$, also known as Chevalier's diagram \citep{Chevalier_1998}, is a useful tool to compare radio observations of different CCSNe. By plotting lines of equal shock velocity and mass-loss rates in this phase space, based on Eq. \ref{eq: Radius_chevalier}, \ref{eq: Magnetic_chevalier}, \ref{eq: b_rho_relation}, and \ref{eq: rho_density_param}, we compare the physical properties of different SNe. In Fig. \ref{fig: Chevalier_diagram} we present all the radio-detected SNe from our sample in a Chevalier's diagram (we use different markers to distinguish between the SNe sub-types). For SNe with a limit on the peak flux density, we use an arrow to point in the direction of the limit. Also marked on this plot are lines of equal shock velocities (using Eq. \ref{eq: Radius_chevalier}), and lines of equal mass-loss rate divided by wind velocity (using Eq. \ref{eq: Magnetic_chevalier}, \ref{eq: b_rho_relation}, and \ref{eq: rho_density_param}), when assuming constant expansion, $p=3$, $\epsilon_{\rm e} = \epsilon_{\rm B} = 0.1$ and $f=0.5$. We note here that the analysis of Type II SNe was done using $p=2.4$, however, since we wish to present the comparison between stripped envelope SNe and Type II SNe we show them on the same plot and choose $p=3$ for reference.

As seen from this analysis, while the shock velocities of stripped-envelope SNe (blue squares for type Ib and Ic, and black plus signs for type IIb SNe) are $\geq 10^4 \, \rm km \, s^{-1}$ and their median is $\sim 3 \times 10^4 \, \rm km \, s^{-1}$ (excluding SNe that are associated with relativistic outflows as discussed in \S\ref{subsec:special_cases}), type II SNe (red circles for regular type II SNe and red diamonds for type IIn SNe) experience significantly slower shocks. The median of type II SNe shock velocities is $\sim 5000 \, \rm km \, s^{-1}$ with the fastest being $\sim 1.1 \times 10^ 4  \, \rm km \, s^{-1}$. At first sight, the SN-CSM shock velocities measured for some type II SNe in our sample may conflict with the photospheric optical velocities which tend to be around $10^4  \, \rm km \, s^{-1}$ for type II SNe. However, some of the SN-CSM shock velocities measured via our radio analysis may represent a late-time velocity, obtained much later after the optical data was obtained, mostly at times $>400$ days after the explosion, and when the ejecta may have decelerated. We also note that FFA can effect the radio emission of some type II SNe with high mass-loss rates and slow wind velocities. In these cases, as the peak frequency and flux density do not depend only on SSA, the mass-loss rate we derived based on this SSA analysis will become lower limits. \citep{Weiler_2002, chevalier_fransson_2017, bietenholz_2020}. We also note here that in the case of a power-law evolution of the radius with time, $R \sim t^{m}$ ($m \leq 1$ for a non-accelerating shockwave), the true shock velocity is smaller, $v_{\rm sh} = m R/t$. As explained in \S\ref{sec: non-detections}, to avoid complicating the analysis by introducing an unknown power-law index, $m$, we limit our analysis of radio upper limits to observations made in the first 18 months after the explosion and assume that the shock has not decelerated significantly during this time (i.e. $m=1$).

Examining the mass-loss rate divided by wind velocity parameter of these radio-detected SNe show that type II (excluding type IIn) SNe exhibit $4 \times 10^{-9} < \frac{\dot{M} \rm{\left[ M_{\odot} \, yr^{-1} \right]}}{v_{\rm w} \rm{\left[ km \, s^{-1} \right]}} < 10^{-5}$) with a median of $\dot{M}/v_{\rm w} \sim 10^{-6} \, \frac{\rm{M_{\odot} \, yr^{-1}}}{\rm{km \, s^{-1}}}$. If, instead, the absorption mechanism is purely FFA, using Eq. \ref{eq: FFA optical depth} and an assumed shock velocity of $10,000 \, \rm km \, s^{-1}$ will increase the median and inferred minimum and maximum mass-loss rate values by about an order of magnitude. Stripped envelope SNe on the other hand exhibit $10^{-10} < \frac{\dot{M} \rm{\left[ M_{\odot} \, yr^{-1} \right]}}{v_{\rm w} \rm{\left[ km \, s^{-1} \right]}} < 5 \times 10^{-6}$ with a median of $\dot{M}/v_{\rm w} = 2 \times 10^{-8} \, \frac{\rm{M_{\odot} \, yr^{-1}}}{\rm{km \, s^{-1}}}$. 

The CSM densities we derived for type II and stripped-envelope SNe are in agreement with the values suggested by stellar evolution models and observational prescription for mass-loss rates \citep{Annu_rev_Smith_2014}. For example, the progenitors of type II-P SNe exhibit mass-loss rates of $10^{-6} - 10^{-5} \rm \, M_{\odot} \, yr^{-1}$ with wind velocities of $10-20 \, \rm km \, s^{-1}$, and those of type II-L SNe are expected to exhibit mass-loss rates of $10^{-5} - 10^{-4} \rm \, M_{\odot} \, yr^{-1}$ with wind velocities of $20-40 \, \rm km \, s^{-1}$. This results in a range of $\dot{M}/v_{\rm w} = 5 \times 10^{-8} - 5 \times 10^{-5} \, \rm  \frac{M_{\odot} \, yr^{-1}}{km \, s^{-1}}$ which is in agreement with the high values we infer for type II SNe. He-stars and Wolf-Rayet stars, which are the suggested progenitors of Ib and Ic SNe, exhibit mass-loss rates of $10^{-7} - 10^{-4} \, \rm M_{\odot} \, yr^{-1}$ wind winds of $1000 \, \rm km \, s^{-1}$ (which translates to $10^{-10} < \frac{\dot{M} \rm{\left[ M_{\odot} \, yr^{-1} \right]}}{v_{\rm w} \rm{\left[ km \, s^{-1} \right]}} < 10^{-7}$). This covers, together with the proposed progenitor of type IIb SNe (with mass-loss rates of $10^{-5} - 10^{-4} \, \rm M_{\odot} \, yr^{-1}$ wind winds of $20-100 \, \rm km \, s^{-1}$ which translates to $10^{-7} < \frac{\dot{M} \rm{\left[ M_{\odot} \, yr^{-1} \right]}}{v_{\rm w} \rm{\left[ km \, s^{-1} \right]}} < 5 \times 10^{-5}$), the entire range we find for stripped-envelope SNe.

This analysis also reflects the high mass-loss rates, of $> {\rm few} \times 10^{-3} \, \rm{M_{\odot} \, yr^{-1}}$ (for wind velocity of $10 \, \rm km \, s^{-1}$), from the progenitors of type IIn SNe. This matches the interpretation of their optical spectra in which the observed strong and narrow emission lines are assumed to be a result of interaction with highly dense CSM. These high densities correspond to high mass-loss rates are not easily explained by mass-loss in stellar winds and may result from short-lived, episodic, mass-loss processes. Episodic mass-loss processes might occur in close binary interaction or eruptions of Luminous Blue Variables (LBV) and the most extreme cool hyper-giants. These are the only classes of stars with observed high wind densities that can form the narrow lines (see the discussions in \citealt{smith_2017} and \citealt{Poonam_2018}, and the possible progenitors of e.g., SN\,2010jl; \citealt{Smith_2011}, and SN\,2005gl; \citealt{Gal_Yam_2007}).

\subsection{Treating special cases in further analysis}
\label{subsec:special_cases}

The analysis presented above showed some SNe that their observational and physical properties deviate from our model assumptions (e.g. relativistic velocities, shock deceleration, and multi-peaked light curves). We briefly discuss these special cases here:
\begin{itemize}
    \item \textbf{SN\,1998bw, SN\,2002ap, SN\,2009bb, and SN\,2012ap} are SNe that are associated with relativistic ejecta and thus need to be treated differently (SN1998bw \citealt{SN1998bw}; SN2002ap \citealt{SN2002ap}; SN2009bb \citealt{SN2009bb} and SN2012ap \citealt{SN2012ap}). They are shown in Fig. \ref{fig: Chevalier_diagram} only for reference, and we exclude these SNe in further analysis.
    
    \item \textbf{SN\,2014C and SN\,2007bg} exhibit several peaks in their radio light curve. This suggests that the SN ejecta interacts with multiple CSM shells deposited in separate mass loss stages. Since the mass-loss history of these SNe clearly diverts from the scenario of constant mass-loss in steady winds we do not use SN\,2014C and SN\,2007bg in our analysis.
    
    \item \textbf{SNe of type IIn} show narrow optical lines that are associated with the interaction of the ejecta with a dense CSM which is not in agreement with the scenario of constant mass-loss by steady winds. This results in a different radio evolution than seen in regular type II (including II-L and IIP) and is therefore marked as a different group. Due to their non-typical radio evolution and mass-loss processes, in the following analysis, we exclude all type IIn SNe.
\end{itemize}
After removing these $16$ cases, when further analyzing radio-detected SNe we are left with $18$ Type II SNe and $38$ stripped-envelope SNe. Out of these radio-detected SNe, the radio spectral peak of $6$ Type II and $12$ stripped-envelope SNe was not observed, and only a lower limit on the peak flux density was given. Therefore, we have an observed peak for $12$ type II SNe, and $26$ stripped envelope SNe.

\section{Constraining the CSM density phase space}
\label{sec: non-detections}

\begin{figure}
\begin{center}
\includegraphics[width=\linewidth]{"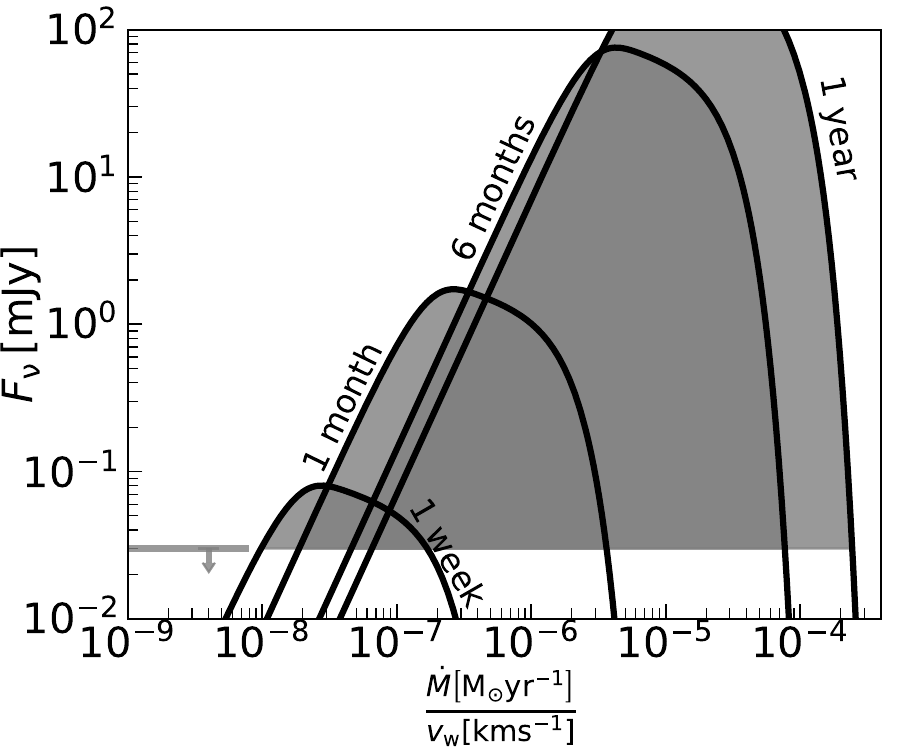"}
\caption{\footnotesize{The radio flux density (at 15.5 GHz) as a function of $\dot{M}/v_{\rm w}$ under the SN-CSM interaction model presented in \S\ref{sec:Modeling} at different timescales (a week, a month, six months, and a year) after the SN explosion. A $3\sigma$ upper limit of $0.03$ mJy (plotted on the bottom left) is translated to ruled out regions in $\dot{M}/v_{\rm w}$ as all values of $\dot{M}/v_{\rm w}$ that produce higher flux densities than the upper limit are ruled out (this is seen in the shaded areas). Here we assumed synchrotron emission from an SN at a distance of $20$ Mpc, shock velocity of $10^4 \, \rm km s^{-1}$, $\epsilon_{\rm e} = \epsilon_{\rm B} = 0.1$, $p=3$, and $f=0.5$, and a temperature of the electrons of $10^{5} \, \rm K$}}
\label{fig: limits_example}
\end{center}
\end{figure}

\begin{figure*}
\begin{center}
\includegraphics[width=0.8\linewidth]{"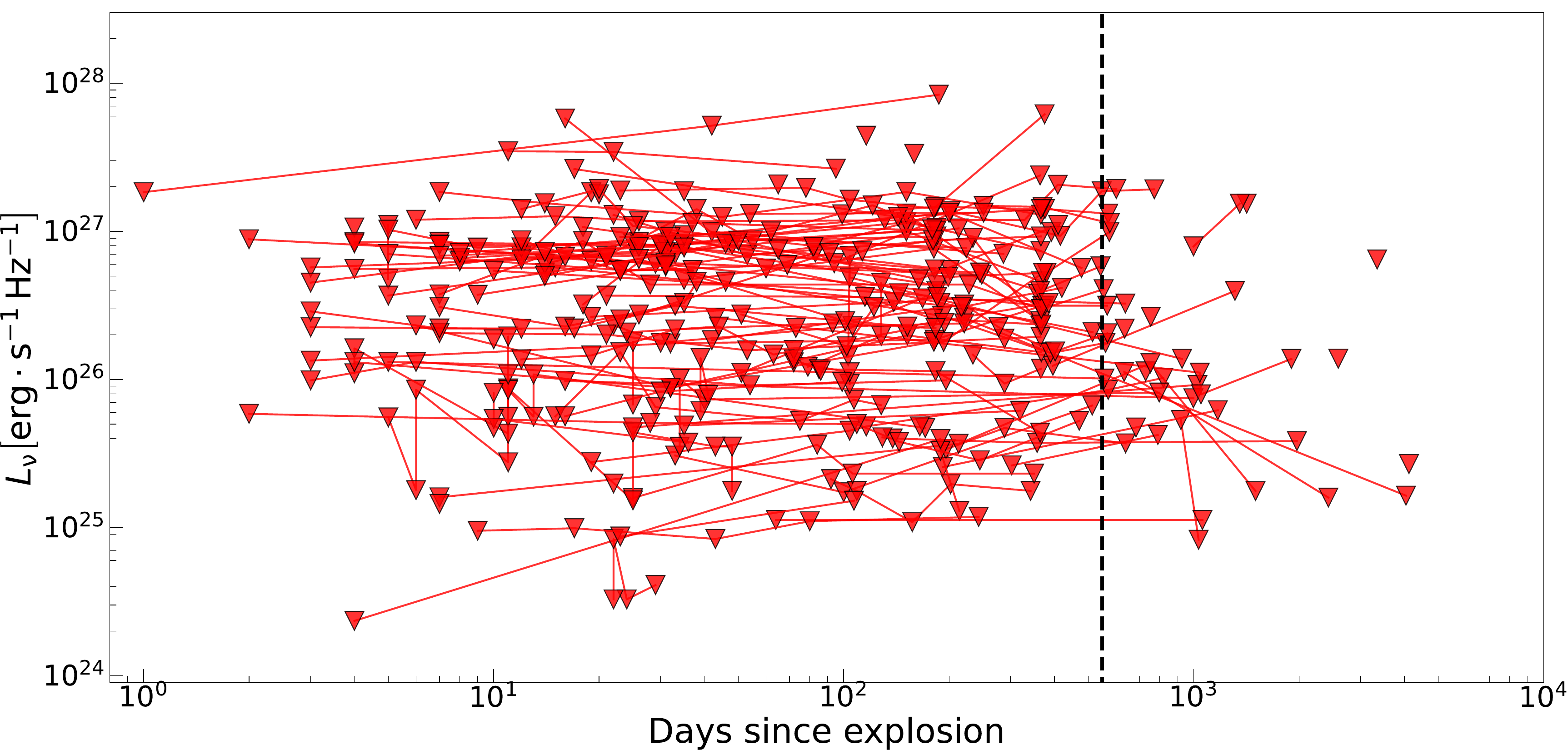"}
\\
\includegraphics[width=0.8\linewidth]{"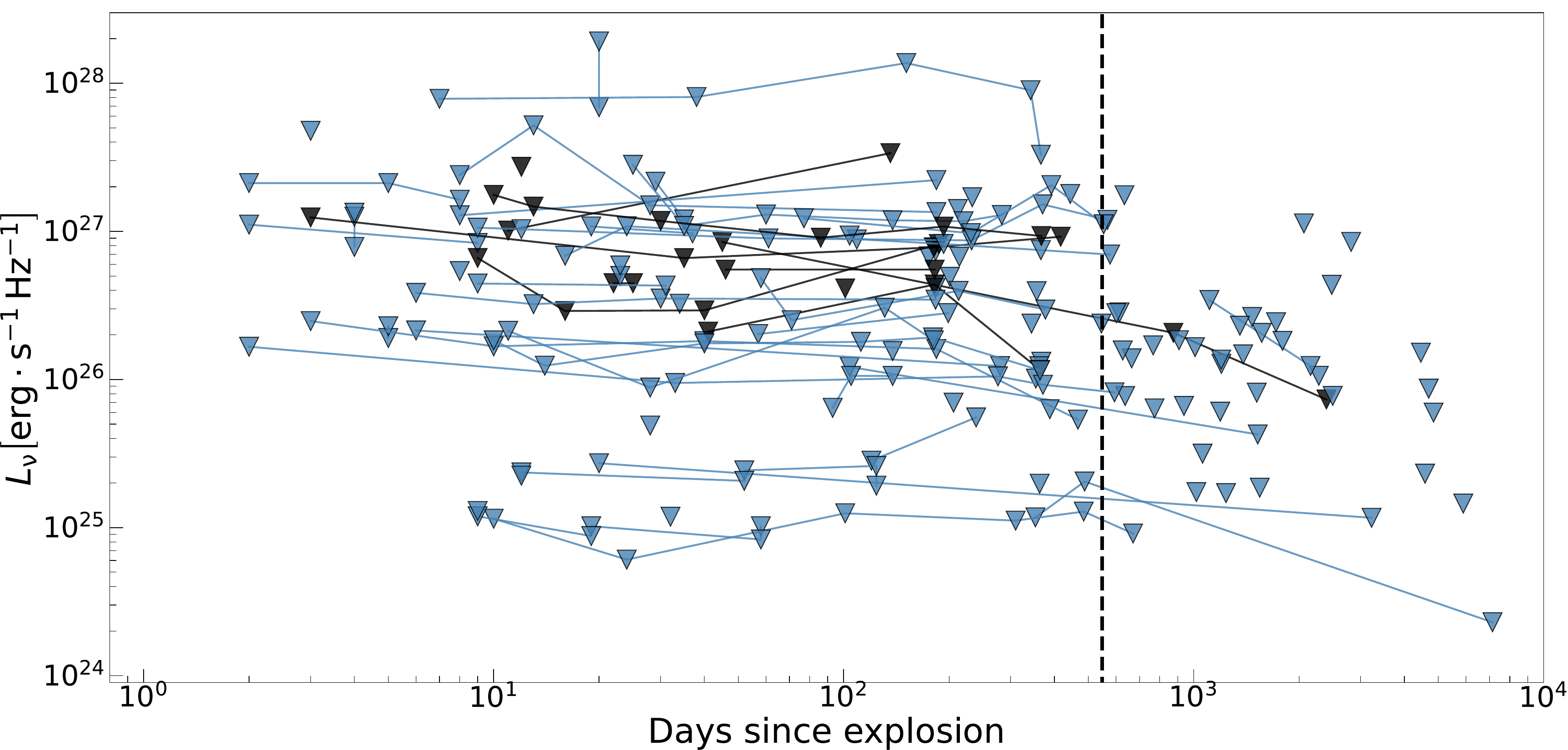"}
\caption{\footnotesize{Limits on the specific radio luminosity from the SNe in our sample that were not detected in radio wavelengths. Each limit is marked with a triangle, the lines are connecting limits from the same SN. The top panel shows the limits on the radio emission from type II SNe (excluding type IIn as discussed in \S\ref{subsec:special_cases}). The bottom panel shows the limits on the radio emission from stripped-envelope SNe (blue is for SNe of type Ib/Ic, and black is for type IIb SNe). Also marked for reference (in a vertical dashed line) is the year-and-a-half time scale as we limit our analysis to flux density upper limits obtained up to a year and a half after the explosion. This is due to the possible shock deceleration at late times as seen in some Type II SNe in \S\ref{sec: detections}}}
\label{fig: all_limits}
\end{center}
\end{figure*}

We now aim to probe the phase space of CSM density around massive stars (and the resulting mass-loss rates) using the full CCSNe sample presented in \S\ref{sec:sample} (radio-detected and non-detected SNe; see a plot of all the radio upper limits in our sample in Fig. \ref{fig: all_limits}) under the same model presented in \S\ref{sec:Modeling}. 

For a given SN with a flux density upper limit, $F_{\rm \nu}$, taken at a frequency, $\nu$, Eq. \ref{eq: full_ssa} (multiplied by $e^{-\tau_{\rm ff}}$ to account for FFA) provide limits on the possible magnetic fields if we assume a radius for the emitting shell. This radius can be calculated under the assumption of a typical shock velocity\footnote{In the following analysis, we assume shock velocities of $10^4 \, \rm km s^{-1}$ for type II SNe and $3 \times 10^4 \, \rm km s^{-1}$ for stripped-envelope SNe, similar to the velocities inferred for radio-detected SNe in \S\ref{sec: detections}.}, $v_{\rm sh}$, and of constant expansion, $R = v_{\rm sh} \Delta t$, where $\Delta t$ is the time of the observation since the SN explosion. Thus, every SN radio upper limit rules out a region in the CSM density phase space (using the relation in Eq. \ref{eq: b_rho_relation} between the magnetic field strength and the density). These ruled-out regions in the phase space of $\rho_{\rm CSM}$ are translated to ruled-out regions in the phase space of $\dot{M}/v_{\rm w}$ using Eq. \ref{eq: rho_density_param}. In Fig. \ref{fig: limits_example} we show an example of how upper limits on the radio emission are translated to ruled-out regions in $\dot{M}/v_{\rm w}$ phase space, ruled-out regions in $\rho_{\rm CSM}$ are derived similarly.

\begin{figure*}
\begin{center}
\includegraphics[width=0.89\linewidth]{"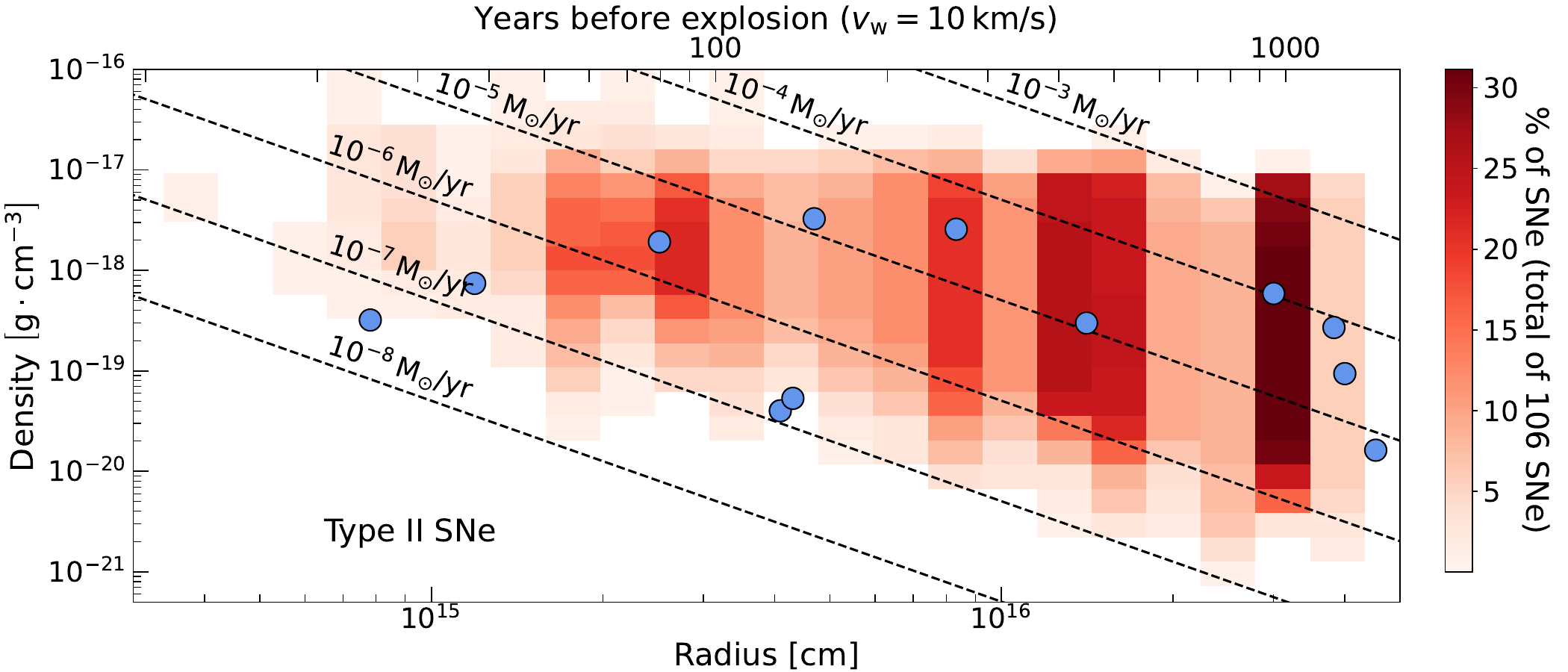"} \\
\smallskip
\smallskip
\smallskip
\smallskip
\includegraphics[width=0.89\linewidth]{"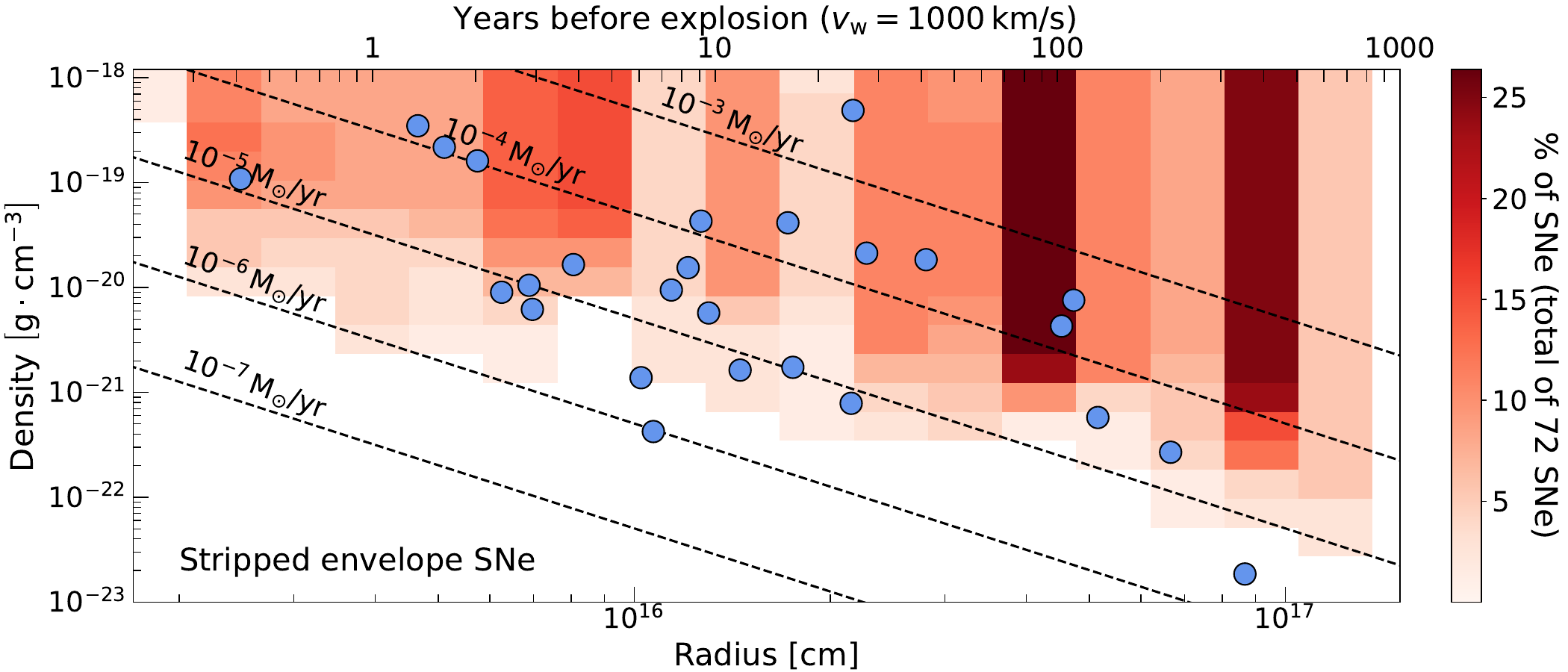"}
\caption{\footnotesize{These plots summarize the ruled out regions in phase space of the density profile around the CCSNe in our sample. The color map shows the percentage of type II (top panel) and stripped-envelope (bottom panel) SNe in our sample that rule out a density for a given radius (both SNe that were not detected in radio wavelengths at all, and SNe with limits on their peak flux density). Here we assumed a constant shock velocity of $10,000 \, \rm km s^{-1}$ for type II SNe, and $30,000 \, \rm km s^{-1}$ for stripped envelope SNe. The blue markers are of the measured densities from SNe in which the peak of their radio light curve or spectra is observed.}}
\label{fig: density_profile}
\end{center}
\end{figure*}

In Fig. \ref{fig: density_profile} we present the results of applying this analysis to the entire CCSNe sample\footnote{We note here that we limit our analysis to flux density upper limits obtained up to a year and a half after the explosion. This is due to the possible shock deceleration at late times. Therefore, when analyzing non-detected SNe we are left with a total of 148 out of 226 (not including the removal of the type IIn SNe mentioned in \S\ref{subsec:special_cases}).}. As seen from this analysis, we managed to probe the density around massive stars to the extent of $\sim 10^{15} - 10^{17}$ cm. For type II SNe this translates to mass-loss at the last $\sim 10-2000$ years of stellar evolution (assuming wind velocity of $10 \, \rm km \, s^{-1}$), and the last few months up to 1000 years of stellar evolution for stripped-envelope SNe. Our systematic observations with AMI-LA play a key role here by ruling out mass-loss rates over the extent of thousands of years before the explosion. According to Fig. \ref{fig: density_profile}, for the progenitors of at least $25\%$ of the type II SNe in our sample we rule-out mass-loss rates in the range of $2 \times 10^{-7} - 10^{-4} \, \rm M_{\odot} \, yr^{-1}$ (assuming wind velocity of $10 \, \rm km \, s^{-1}$). For the progenitors of at least $20\%$ of the stripped-envelope SNe in our sample (as suggested by Fig. \ref{fig: density_profile}) we rule-out mass-loss rates $\gtrsim 5 \times 10^{-5} \, \rm M_{\odot} \, yr^{-1}$ (assuming wind velocity of $1000 \, \rm km \, s^{-1}$). 

\begin{figure*}
\begin{center}
\includegraphics[width=\linewidth]{"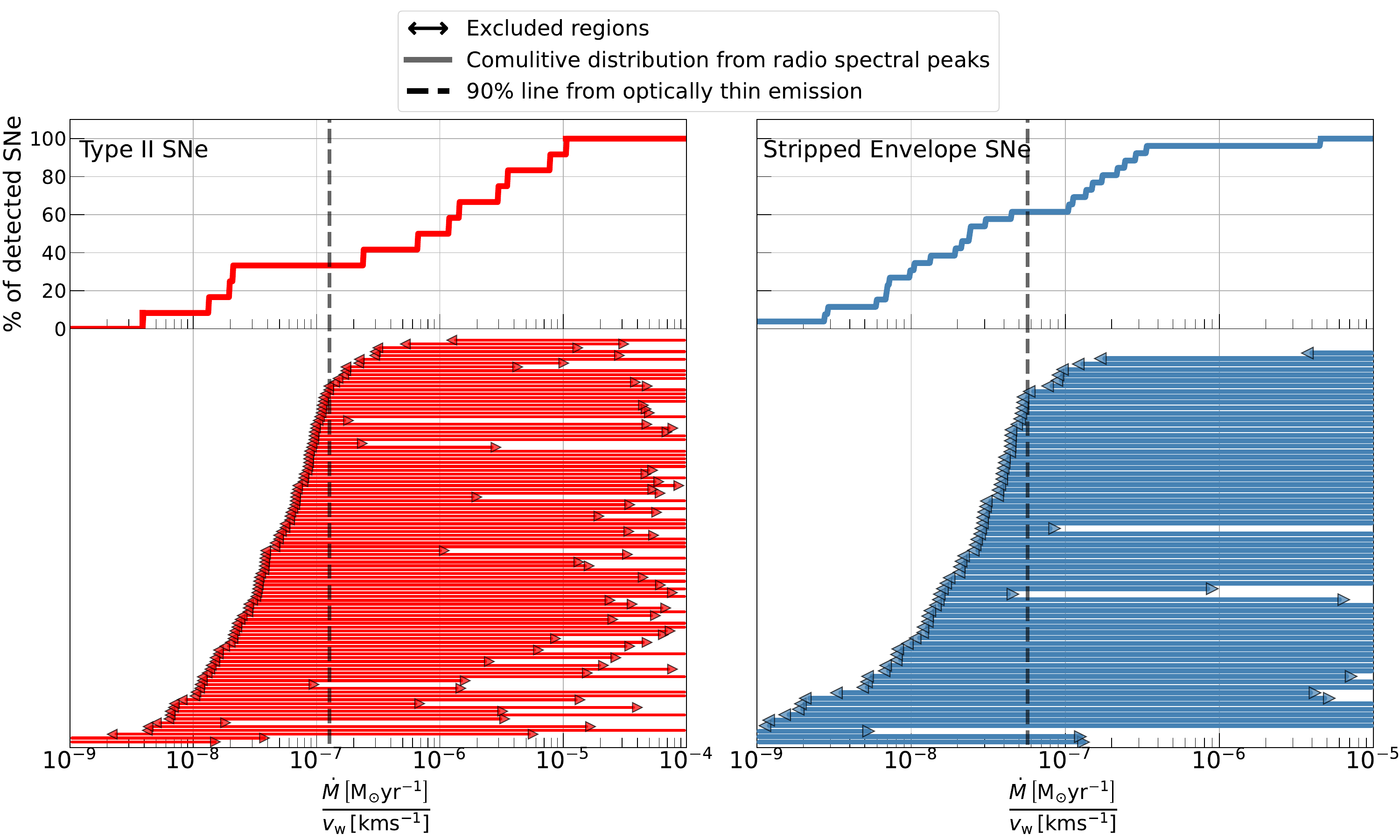"}
\caption{\footnotesize{Representation of the phase space of mass-loss rate divided by wind velocity using combined data from both radio-detected and radio-non-detected SNe. This plot shows the ruled-out regions as limits on $\dot{M}/v_{\rm w}$ (as described in \S\ref{sec: non-detections}), and the distribution of this parameter as obtained from a peak in the radio light curve or spectra of radio-detected SNe (as described in section \ref{sec: detections}). \textbf{Left panel} shows the results for Type II SNe while the \textbf{right panel} is for stripped-envelope SNe. Shock velocities of $10^4$\,km/s are assumed for Type II SNe while for stripped-envelope SNe we assumed $3 \times 10^4$\,km/s. In this plot, there are $56$ radio detected ($18$ type II, and $38$ stripped-envelope SNe) and $160$ radio-non-detected SNe ($100$ type II, and $60$ stripped-envelope SNe). The radio peak has been observed for $38$ of them ($12$ type II, and $26$ stripped-envelope SNe) and is represented in the histograms, and for $18$ we only have limits on their spectral peak, represented in the ruled-out regions. Also plotted for reference are the lines that correspond to the $90 \%$ of the SNe with limits on the CSM phase space which rule out values of $\dot{M}/v_{\rm w}$ greater than it. In Fig. \ref{fig: Forbidden regions combined - mass loss} in appendix \ref{sec: mass_loss_plots_appendix} we show the same results for typical wind velocities, i.e., the phase space of mass-loss rate (not divided by wind velocity).}
\label{fig: Forbidden regions combined}}
\end{center}
\end{figure*} 

\begin{figure*}
\begin{center}
\includegraphics[width=0.495\linewidth]{"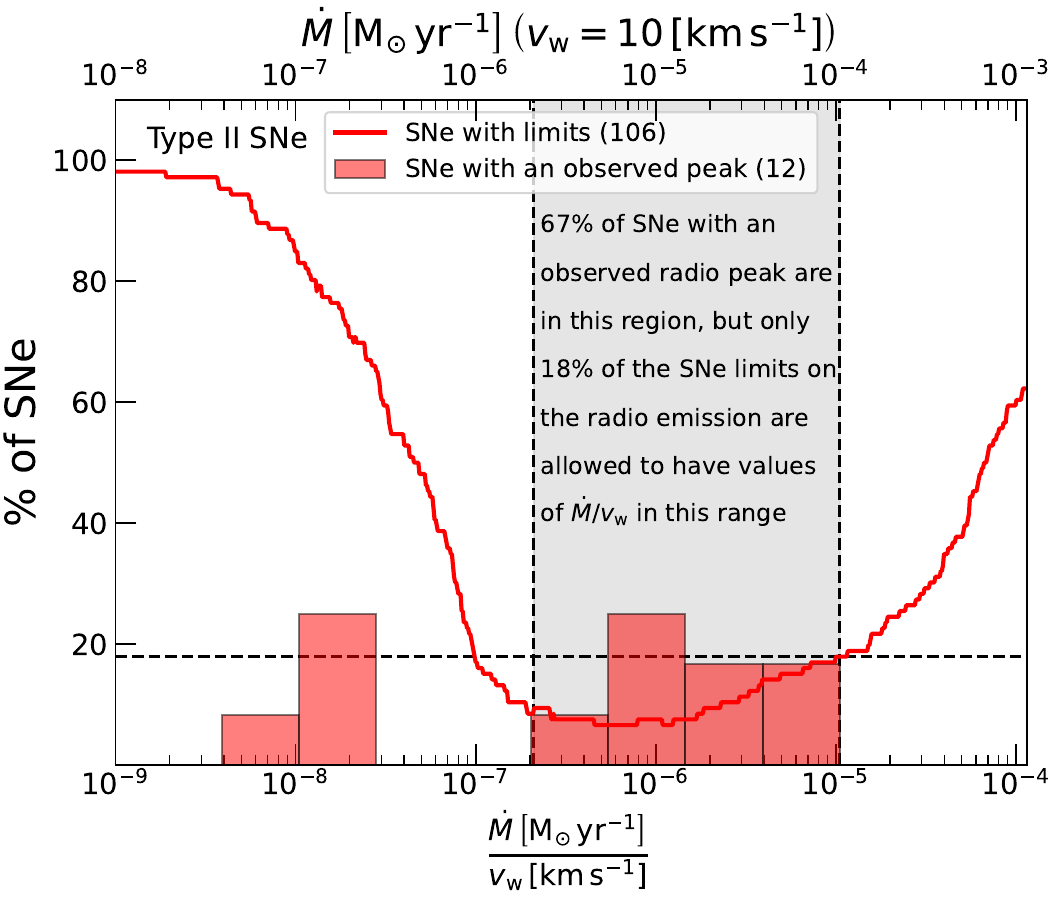"}
\includegraphics[width=0.495\linewidth]{"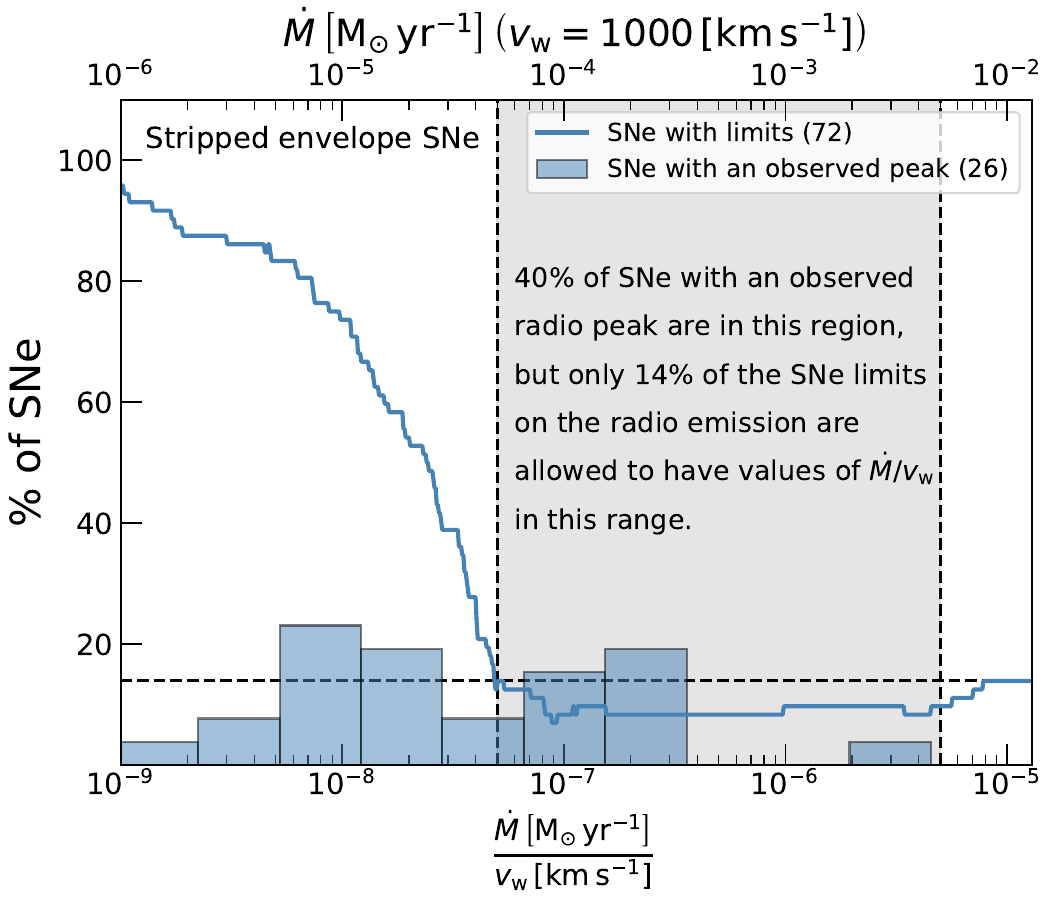"}
\caption{\footnotesize{Comparison of mass-loss history between SNe with an observed radio peak, and SNe with radio limits. For the SNe with an observed radio peak we present the distribution of $\dot{M}/v_{\rm w}$ (bar histogram). For the SNe with limits on their radio emission, or on their radio peak, we present the percentages of SNe that are allowed for each value of $\dot{M}/v_{\rm w}$ (step diagram). The left panel shows the results for type II SNe and the right panel is for stripped-envelope SNe. Also marked in shaded regions is the area in the phase space of $\dot{M}/v_{\rm w}$ that shows the discrepancy between the values inferred from SNe with radio peaks and SNe with limits. We also evaluate this phase space for mass-loss rates assuming typical wind velocities (top x-axes).}
\label{fig: mass_loss_dist}}
\end{center}
\end{figure*}

While it is useful to probe the ruled-out density profiles seen in Fig. \ref{fig: density_profile}, another way to look at these limits is by probing the ruled-out regions in the phase space of mass-loss rate divided by wind velocity. Since each SN might have several limits, stacking them will give a range of ruled-out $\dot{M}/v_{\rm w}$ for each SN. Here we assumed that the CSM was deposited via constant mass-loss in steady winds throughout the entire evolution of the progenitor star. We show this analysis in Fig. \ref{fig: Forbidden regions combined} (In Fig. \ref{fig: Forbidden regions combined - mass loss} in appendix \ref{sec: mass_loss_plots_appendix} we show the same results for typical wind velocities), and a comparison in percentages in Fig. \ref{fig: mass_loss_dist}. As suggested in the left panels of these figures, assuming $v_{\rm w}=10 \, \rm km \, s^{-1}$, for $82 \%$ (87 out of 106) of the Type II SNe with limits on their mass-loss rate, the region of mass-loss rate between $2 \times 10^{-6}$ and $10^{-4}\, \rm M_{\odot} \, yr^{-1}$ is ruled out. Values of mass-loss rates in that range are inferred for $67 \%$ (8 out of 12) of the SNe with an observed peak in the radio. In the case of stripped-envelope SNe (right panels of Figures \ref{fig: Forbidden regions combined} and \ref{fig: mass_loss_dist}), assuming $v_{\rm w} = 1000 \, \rm km \, s^{-1}$ suggests that $\sim 86 \%$ (62 out of 72) of the SNe with limits on their mass-loss rate rules out the region of $5 \times 10^{-5} - 5 \times 10^{-3} \, \rm M_{\odot} \, yr^{-1}$. A value of mass-loss rate in this region is inferred for $\sim 40 \%$ (11 out of 26) of the stripped-envelope SNe with an observed radio peak. 

This analysis points to a discrepancy between the values of $\dot{M}/v_{\rm w}$ inferred from radio-detected SNe and radio-non-detected SNe, especially for the case of Type II SNe. Thus, detecting SNe in radio wavelengths is highly biased towards progenitor stars that experience high mass-loss rates at their last stages of stellar evolution. As noted above, if we change our assumption to include FFA as the dominant absorption mechanism for type II SNe with an observed radio peak, the range of mass-loss rate values will increase by about an order of magnitude. However, the discrepancy between radio-detected and radio-non-detected SNe still remains. If the peak of the radio emission of type II SNe is dominated by FFA the range of $\sim 10^{-7}$ and $\sim 3 \times 10^{-4}\, \rm M_{\odot} \, yr^{-1}$ is inferred for $67\%$ of the SNe with an observed radio peak (for an assumed $v_{\rm w} = 10 \, \rm km \, s^{-1}$). This range is ruled out by $70\%$ of the SNe with limits on their radio emission.

Theoretical and observational mass-loss prescriptions predict a range of possible mass-loss rates and wind velocities \citep{Annu_rev_Smith_2014}. A large fraction of these regions are ruled out by large fractions of radio non-detected Type II SNe. For example, the suggested range of $\dot{M}/v_{\rm w}$ for the likely progenitors of Type II-P SNe (RSG) is $5\times 10^{-8} \leq \, \frac{\dot{M} \left[{\rm M_{\odot} \, yr^{-1}} \right]}{v_{\rm w} \left[{\rm km \, s^{-1}} \right]} \leq 10^{-6}$. It has also been suggested that the likely progenitors of Type II-L SNe (RSG/YSG) experience mass-loss rate to wind velocity ratio between $\rm 2.5 \times 10^{-7}$ and $\rm 5 \times 10^{-5} \, \frac{M_{\odot} \, yr^{-1}}{km \, s^{-1}}$. However, we find that the entire range suggested for the progenitors of II-L and II-P SNe is being ruled out by $30 \%$ and $50 \%$, respectively, of the Type II SNe in our sample that only have limits on their mass-loss rates. High values of mass-loss rates ($\gtrsim {\rm few} \times 10^{-3} \, \rm M_{\odot} \, yr^{-1}$ with winds of $10 \, \rm km \, s^{-1}$) cannot be ruled out in this analysis as FFA plays an important role for such cases, especially at early times. However, SNe with such high mass-loss rates (and resulting high densities) should exhibit narrow emission lines in their optical spectrum. We note that we excluded type IIn SNe from our analysis, although we do not have full optical spectral cover for all SNe in our sample.

In this work we provide multi-epoch systematic monitoring on a logarithmic scale of a large sample CCSNe. Systematic monitoring at this scale has never been done before, and these results emphasize the importance of systematic observations and continuous follow-up even when a target is not detected. We also note that while the ranges of mass-loss rates we infer are not different from previous works, in this work we take different approach in our analysis. Past studies often focused on the mass-loss rates from small samples of radio-detected SNe \citep{Chevalier_1998, Weiler_2002}, very late-time observations (years after the explosion) corresponding to earlier epochs of stellar evolution compared to the SN explosion \citep{rose_2024}, and average values and width of the distributions of mass-loss rates based on a large sample of radio-detected and non-detected CCSNe \citep{bietenholz_2020}. We, on the other hand, use our sample to probe the phase space of mass-loss rate and to compare the distribution of mass-loss rates inferred from radio-detected SNe to the ruled-out regions due to radio upper limits. This comparison leads to the conclusion that there is a large discrepancy between the progenitors of radio-detected and non-detected SN regarding mass loss during their final stages of evolution.

\section{Testing model assumptions}
\label{sec: caveats}

\begin{figure*}
\begin{center}
\includegraphics[width=0.48\linewidth]{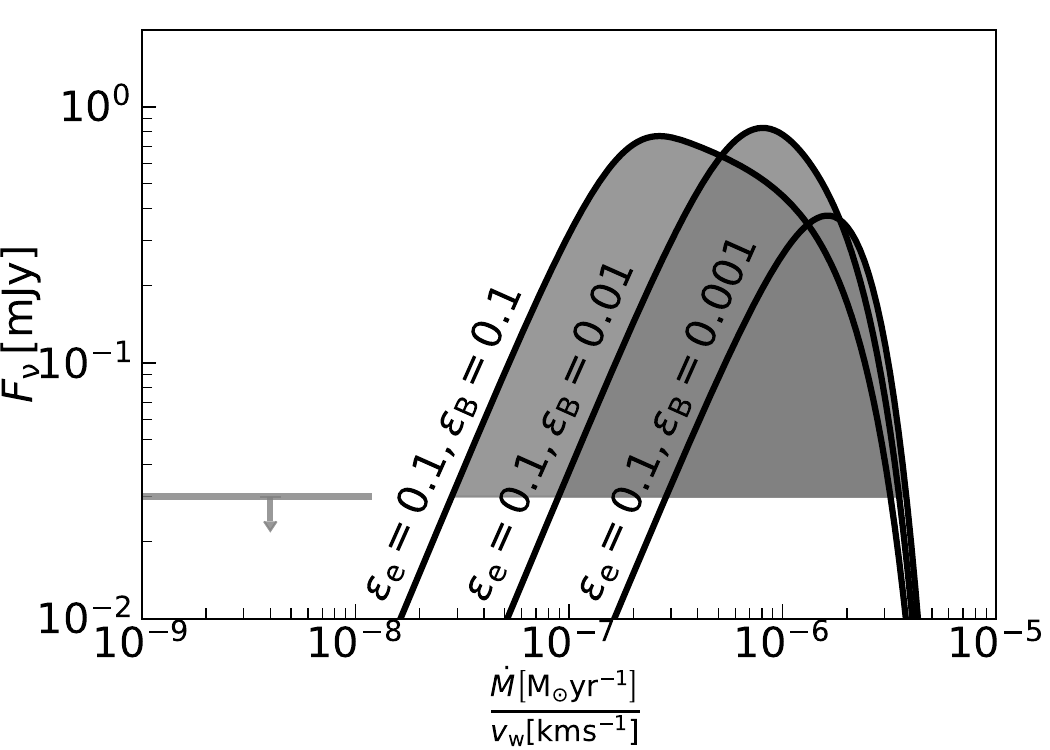}
\smallskip
\includegraphics[width=0.48\linewidth]{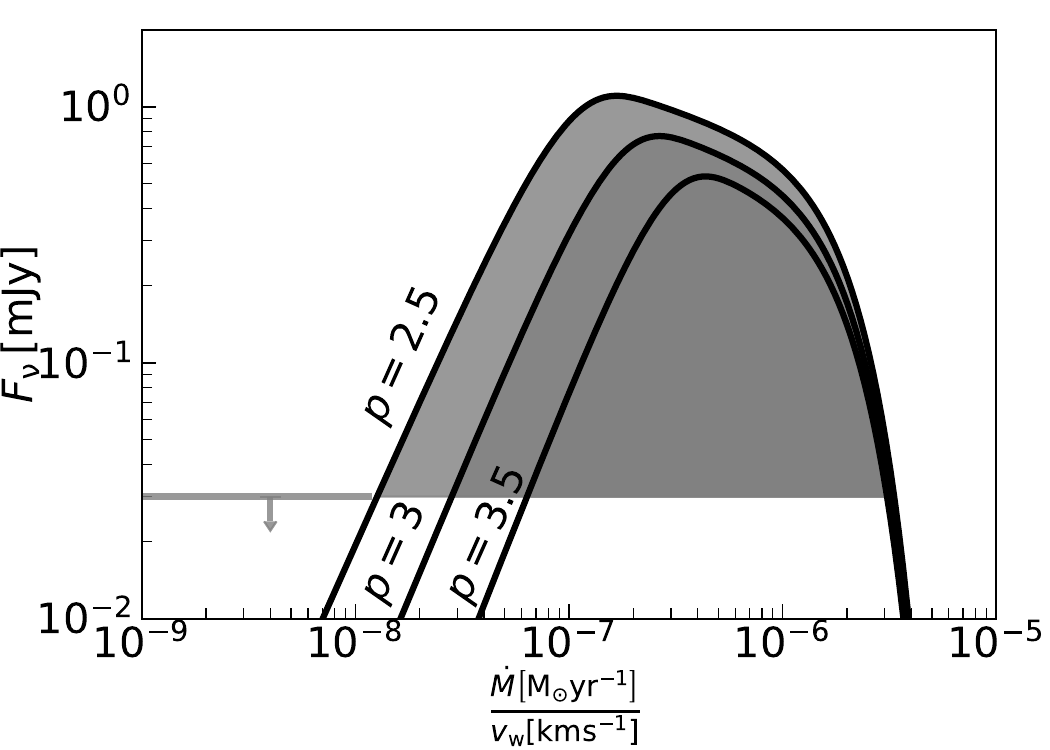}
\caption{\footnotesize{The radio flux density (at 15.5 GHz) as a function of $\dot{M}/v_{\rm w}$ under the SN-CSM interaction model presented in \S\ref{sec:Modeling} at 30 days after the SN explosion. A $3\sigma$ upper limit of $0.03$ mJy (plotted on the bottom left) is translated to ruled out regions in $\dot{M}/v_{\rm w}$ as all values of $\dot{M}/v_{\rm w}$ that produce flux density higher than the upper limit are ruled out (this is visible through the shaded areas). Here we assumed synchrotron emission from an SN at a distance of $30$ Mpc, and shock velocity of $10^4 \, \rm km s^{-1}$. In the left panel we assume $\epsilon_{\rm e} = 0.1$, $p=3$, $T_{\rm e} = 10^{5} \, \rm K$, and $f=0.5$, while varying $\epsilon_{\rm B}$. In the right panel we assume $\epsilon_{\rm e} = \epsilon_{\rm B} = 0.1$, $T_{\rm e} = 10^{5} \, \rm K$, and $f=0.5$, while varying $p$.}}
\label{fig:testing_assumptions}
\end{center}
\end{figure*}

The analysis presented above, both when we had radio detections (\S\ref{sec: detections}), and when we had only radio upper limits (\S\ref{sec: non-detections}), was performed under several simplifying assumptions. Changing our assumptions, such as the electron energy power-law index, $p$, and energy equipartition ($\epsilon_e = \epsilon_B = 0.1$), will change the derived mass-loss rate. In the following section we test how varying these assumptions will impact our conclusions on the mass-loss rate phase space.

\subsection{Deviation from equipartition}
\label{subsec:equipartition}

Deviations from equipartition has been observed for a handful of SNe (mostly stripped-envelope SNe, e.g., SN\,2011dh; \citealt{SN2011dh_3}, \citealt{SN2011dh_2}, SN\,2012aw; \citealt{SN2012aw}, SN\,2020oi; \citealt{SN2020oi_horesh}), which all point towards $f_{\rm eB} > 1$. A typical assumption is that $\epsilon_{\rm e} = 0.1$. Thus, we now discuss deviations where $\epsilon_{\rm e} = 0.1 > \epsilon_B$.

As seen from Eq. \ref{eq: Radius_chevalier} the derivation of the radius of the emitting shell from the radio spectral peak goes as $f_{\rm eB}^{-1/19}$, and therefore, so does the shock velocity (assuming free expansion). For $f_{\rm eB} = 10$, the inferred shock velocity will be reduced by $\sim 11 \%$, and for $f_{\rm eB} = 100$, by $\sim 23 \%$. The magnetic field strength (derived from the radio spectral peak; Eq. \ref{eq: Magnetic_chevalier}) scales as $f_{\rm eB} ^{-4/19}$ and therefore will be reduced by $38 \%$ and $62$ for $f_{\rm eB} = 10$ and $100$, respectively. Overall, changing $f_{\rm eB}$ from $1$ to $10$ and $100$ (assuming $\epsilon_{\rm e} = 0.1$) will increase the mass-loss rate by a factor of $3.8$, and $14$, respectively.

A flux density upper limit will result in a weaker upper limit on the mass-loss rate due to deviation from equipartition (mainly by decreasing $\epsilon_{\rm B}$, see top right panel of Fig. \ref{fig:testing_assumptions}). Therefore, radio non-detection of non-equipartition shock waves traveling in a CSM will be less constraining on the low end of the mass-loss rate parameter space (for reference, we show the results of this analysis assuming $\epsilon_{\rm e} = 0.1$ and $\epsilon_{\rm B} = 0.01$ in \ref{sec: different_parameters_appendix}).

\subsection{Electron energy power-law index}
\label{subsec:powerlaw}

Throughout our analysis, we assumed that the accelerated electrons in the shock front gain energy density distribution of $E^{-p}$ with $p=3$ for stripped-envelope SNe and $p=2.4$ for type II SNe. While this is a reasonable assumption, different values of optically thin power-law indices have been observed implying that the values of $p$ are sometimes lower or higher than these values (see Table 1 in \citealt{Chevalier_1998} and Table 2 in \citealt{Weiler_2002} for different power-law indices for the energy distribution of the electrons). Testing the effect of $\Delta p = \pm 0.5$ shows that the shock velocity and the $\dot{M}/v_{\rm w}$ measured from a radio spectral peak (assuming SSA) will change by $17 \%$ and $50 \%$, respectively. We also test the effect on the phase space of ruled out $\dot{M}/v_{\rm w}$ from radio non-detected SNe, by changing $p$. The bottom panel of Fig. \ref{fig:testing_assumptions} shows the radio flux density from an SN as a function of the CSM density parameter, for different values of $p$. A flux density upper limit will rule out smaller regions in the mass-loss rate phase space if we assume $p>3$, and larger regions if we assume $p<3$ (for reference, we show the results of this analysis for type II SNe assuming $p = 2.1$ and $3$ in \ref{sec: different_parameters_appendix}).

\section{Summary and conclusions}
\label{sec:summary}

In this work we analyzed radio observations of CCSNe from a population point of view. We form a sample of more than 300 CCSNe with radio observations, comprised of both archival and new (first presented here) observations. We find that while it is not rare to detect them in radio wavelengths, the majority of CCSNe, about $78 \%$, are not detected in radio wavelengths even when monitored over a large range of time scales. We analyzed the entire sample using the SN-CSM interaction model and probed the resulting phase spaces of CSM density around CCSNe and mass-loss rate from the progenitor massive stars.

Our analysis suggests that most SNe have different values of CSM density around them than what is suggested by only analyzing SNe in which a peak in their radio light curve or spectra is observed. This means that the majority of SNe experience different values of mass-loss rate than typically reported. For example, while $\sim 70 \%$ of the type II SNe with an observed peak in the light curve or spectrum experience $2 \times 10^{-6} \leq \frac{\dot{M}}{\rm M_{\odot} \, yr^{-1}} \leq 10^{-4}$ for an assumed wind of $10 \, \rm km \, s^{-1}$, the vast majority of the other subset of type II SNe (about $\sim 80 \%$) rule out this range of mass-loss rates. For stripped-envelope SNe on the other hand, $40 \%$ of the SNe with an observed peak in their radio light curve or spectra experience $5 \times 10^{-5} \leq \frac{\dot{M}}{\rm M_{\odot} \, yr^{-1}} \leq 5 \times 10^{-3}$ for an assumed wind of $1000 \, \rm km \, s^{-1}$. However, $86 \%$ of the stripped-envelope SNe with limits on their mass-loss rate point to $\dot{M} \leq 5 \times 10^{-5} \, \rm M_{\odot} \, yr^{-1}$.

Red supergiants (RSGs) are the likely progenitors of Type II-P and II-L SNe, which are the majority of Type II SNe in our sample. These progenitors can experience mass-loss rates between $10^{-6}$ to $10^{-5} \, \rm M_{\odot} yr^{-1}$ and wind velocities as high as $40 \, \rm km s^{-1}$ \citep{Annu_rev_Smith_2014}. The low end of this range can explain some of the Type II SNe that rule out values $\leq 4 \times 10^{-6} \, \rm M_{\odot} yr^{-1}$ but there is still a substantial fraction of them that cannot be explained by these models. Furthermore, various stellar evolution models and observations suggest a range of possible mass-loss rates. However, most of it (namely $10^{-6} < \dot{M} < 10^{-4} \, \rm M_{\odot} \, yr^{-1}$ for winds of $10 \, \rm km \, s^{-1}$) is ruled out by the majority of Type II SNe in our sample. 

While the assumption of constant mass-loss rate in steady winds plays a key role in our analysis and conclusions of the phase space of mass-loss rates from massive stars, there are examples of clear deviation from such a simple scenario (see e.g., SN\,2003L; \citealt{SN2003L}, SN\,2014C; \citealt{SN2014C_2}, SN\,2004c; \citealt{SN2004cc_demarchi}, SN\,2019oys; \citealt{SN2019oys}), and other mass-loss mechanisms can take place (e.g., binary stars interactions and violent mass ejections). We emphasize here that this work takes a step forward in constraining the wind model from a statistical point of view. Any future work combining the systematic approach taken here with deeper limits and even more epochs, constraining even larger radii around the SN and shorter time-scales compared to the time since the explosion, can shift our understanding of the mass-loss mechanisms that shape massive stars at the end of their lives and possibly determine the role of winds in the last thousand years of evolution.

Our systematic approach, of probing newly reported CCSNe on different timescales after the SN explosion has proven to be valuable. The analysis above amplifies the importance of systematically observing CCSNe in radio wavelengths. More sensitive observations with state-of-the-art facilities (such as the Karl G. Jansky Very Large Array) can improve our understanding of the last stages of stellar evolution by probing the phase space of mass-loss rates to lower values (with early observations) and to higher values (with late-time observations). Furthermore, high cadence observations (on time scales of days) can probe variable mass-loss on time scales of only a few years before the SN explosion.

\section*{Acknowledgments}
\label{sec:ACK}
A.H. is grateful for the support by the Israel Science Foundation (ISF grant 1679/23) and by the United States-Israel Binational Science Foundation (BSF grant 2020203). A.H is supported by the Sir Zelman Cowen Universities Fund. We acknowledge the staff who operate and run the AMI-LA telescope at Lord's Bridge, Cambridge, for the AMI-LA radio data. AMI is supported by the Universities of Cambridge and Oxford, and by the European Research Council under grant ERC-2012-StG-307215 LODESTONE.

\appendix
\twocolumngrid
\section{Different wind velocities}
\label{sec: mass_loss_plots_appendix}

\begin{figure*}
\begin{center}
\includegraphics[width=0.4\linewidth]{"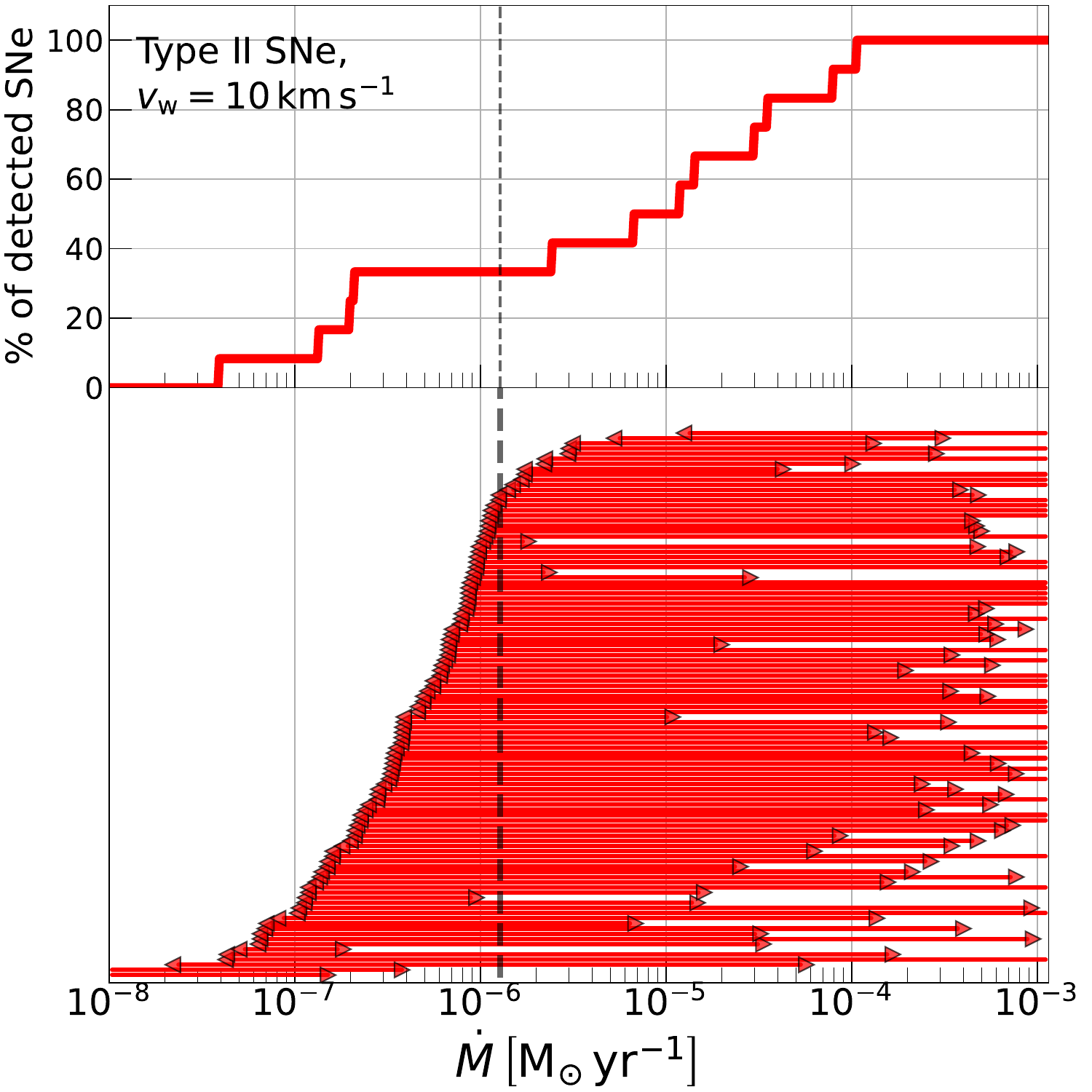"}
\includegraphics[width=0.4\linewidth]{"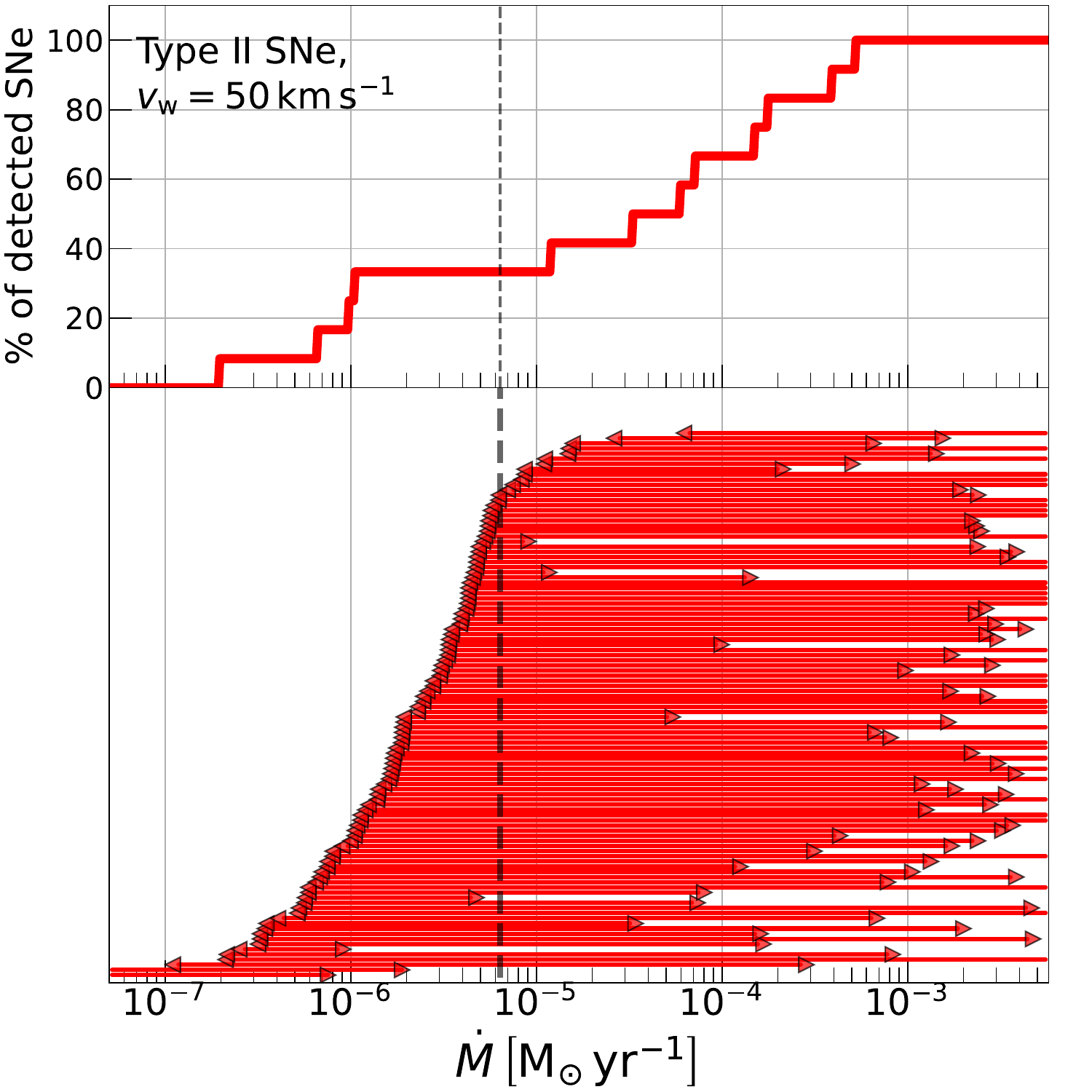"}
\includegraphics[width=0.4\linewidth]{"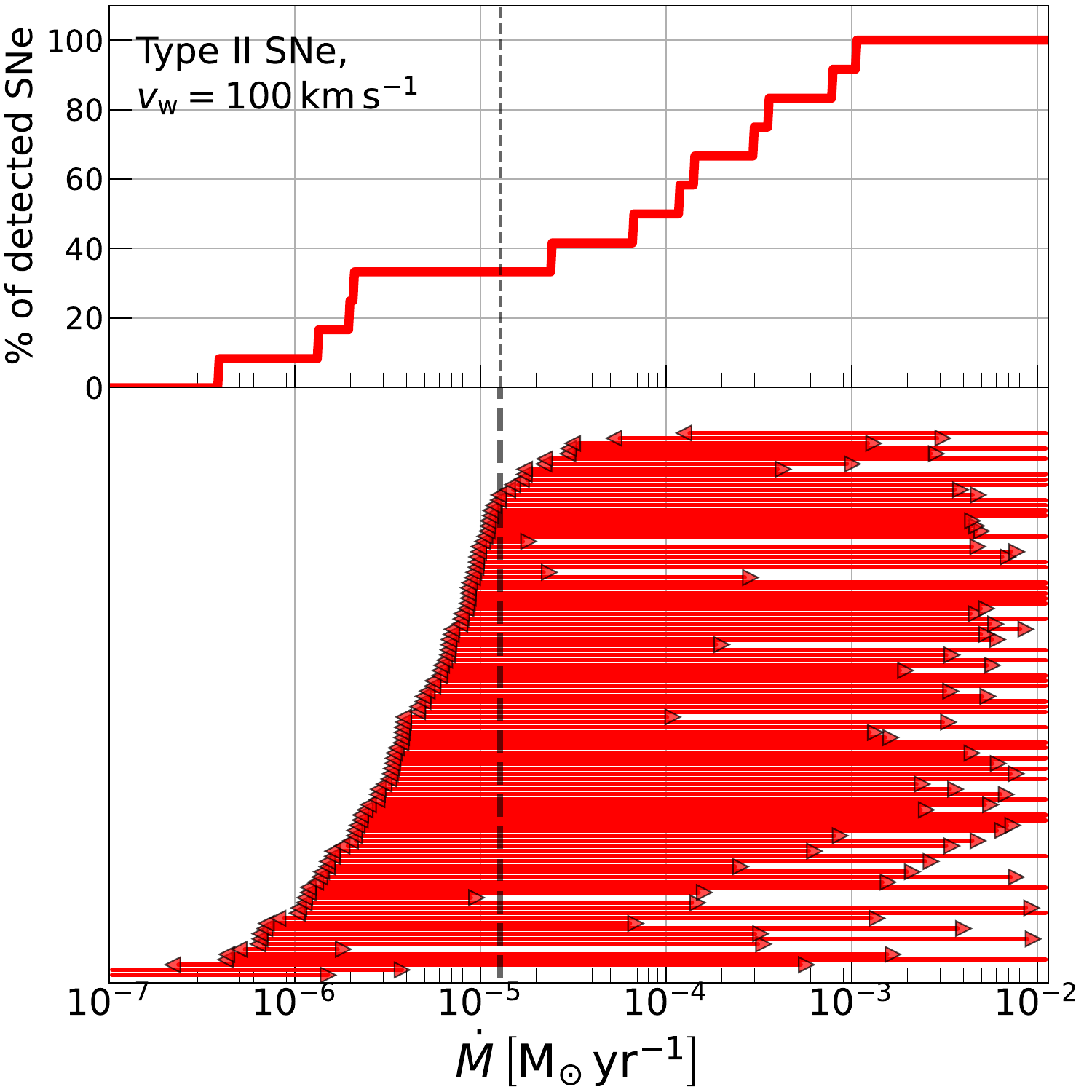"}
\includegraphics[width=0.4\linewidth]{"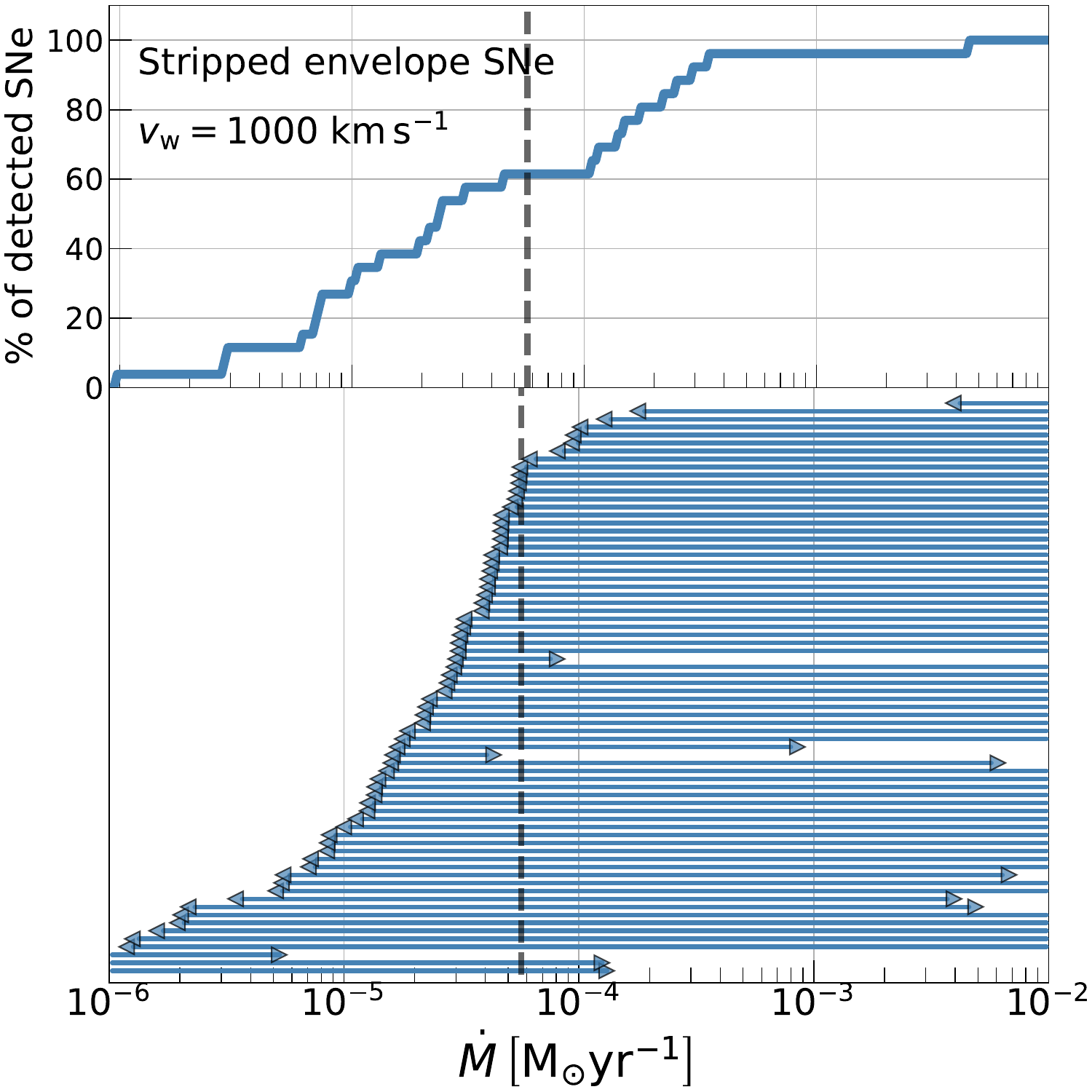"}
\caption{\footnotesize{Representation of the phase space of mass-loss rate for different wind velocities using combined data from both radio-detected and radio-non-detected SNe. These plots are similar to the plots seen in Fig. \ref{fig: Forbidden regions combined} but the values of $\dot{M}/v_{\rm w}$ are multiplied by typical wind velocities. For type II SNe we show the results for $v_{\rm w} = 10 \, \rm km \, s^{-1}$ (top left), $50 \, \rm km \, s^{-1}$ (top right), and $100 \, \rm km \, s^{-1}$ (bottom left). For stripped-envelope SNe we show the results for $v_{\rm w} = 1000 \, \rm km \, s^{-1}$ (bottom right).}
\label{fig: Forbidden regions combined - mass loss}}
\end{center}
\end{figure*}

Our analysis in \S\ref{sec: non-detections} shows the results of the mass-loss rate divided by wind velocity ($\dot{M}/v_{\rm w}$) phase space. However, it is useful to show these results in the context of mass-loss rate under typical wind velocities. In Fig. \ref{fig: Forbidden regions combined - mass loss} we present the phase space of mass-loss rate under wind velocities of $10$, $50$, and $100 \, \rm km \, s^{-1}$ for type II SNe (top left, top right, and bottom left panels), and $1000 \, \rm km \, s^{-1}$ for stripped-envelope SNe (bottom right panel).

\section{Different model parameters}
\label{sec: different_parameters_appendix}

\begin{figure*}
\begin{center}
\includegraphics[width=0.43\linewidth]{"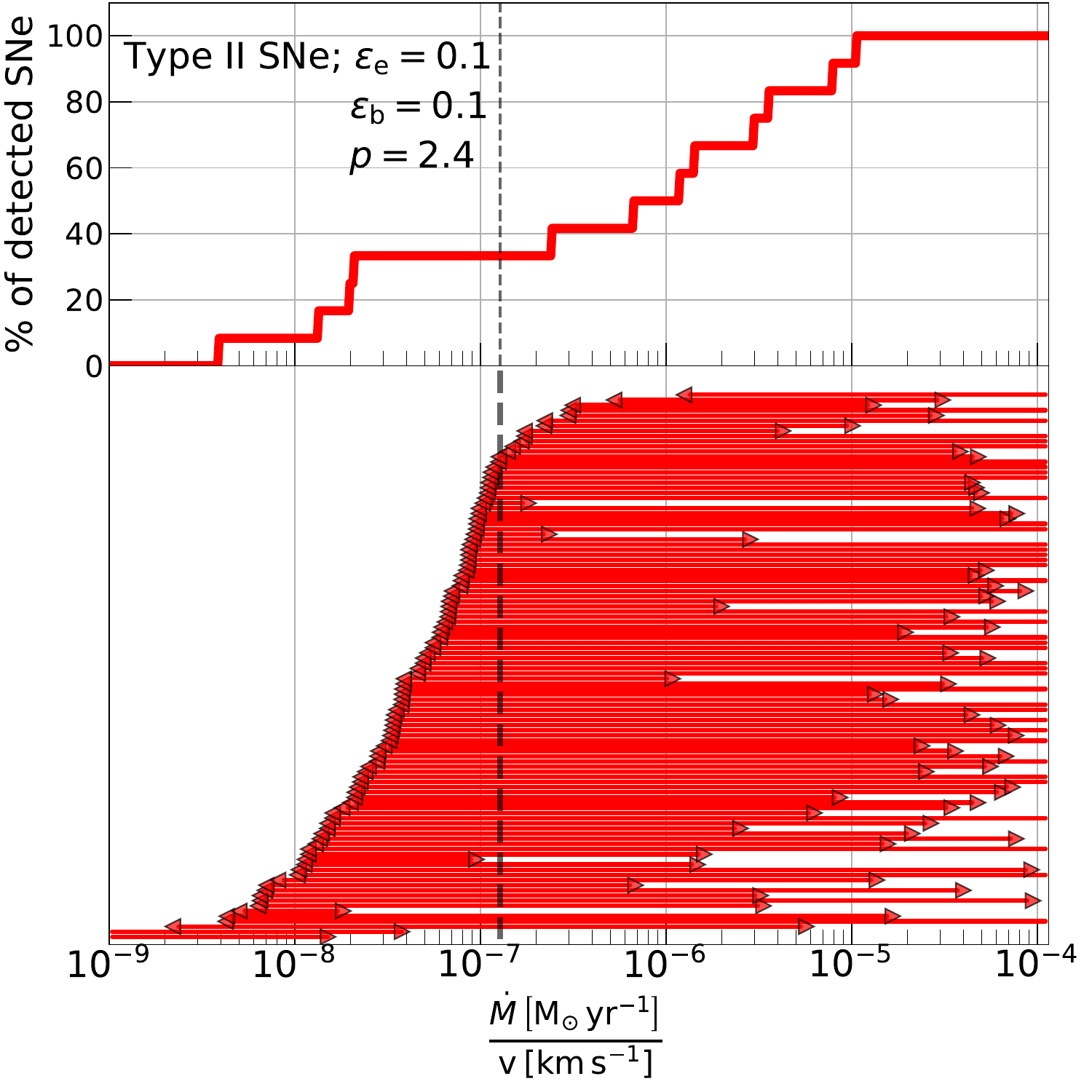"}
\includegraphics[width=0.43\linewidth]{"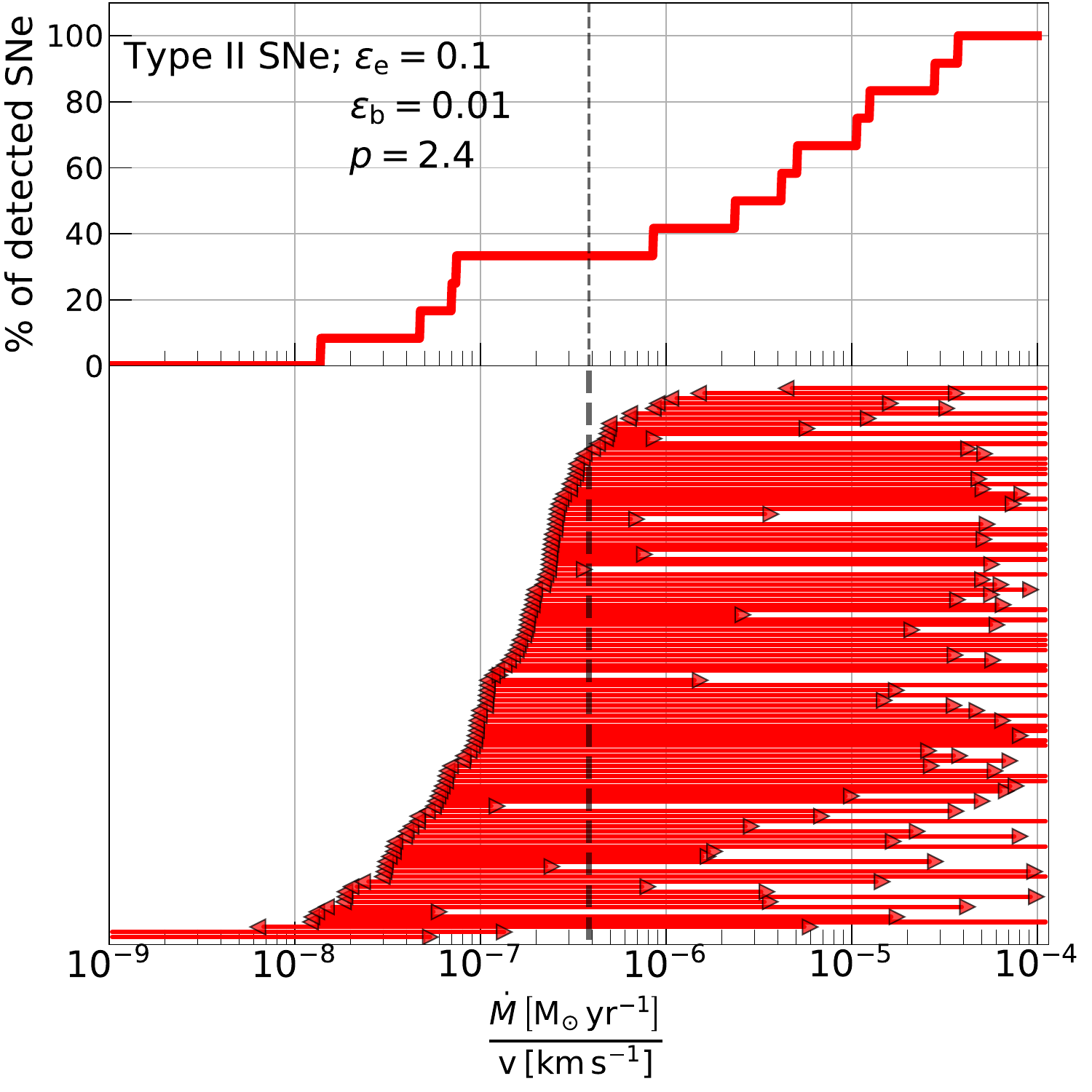"}
\includegraphics[width=0.43\linewidth]{"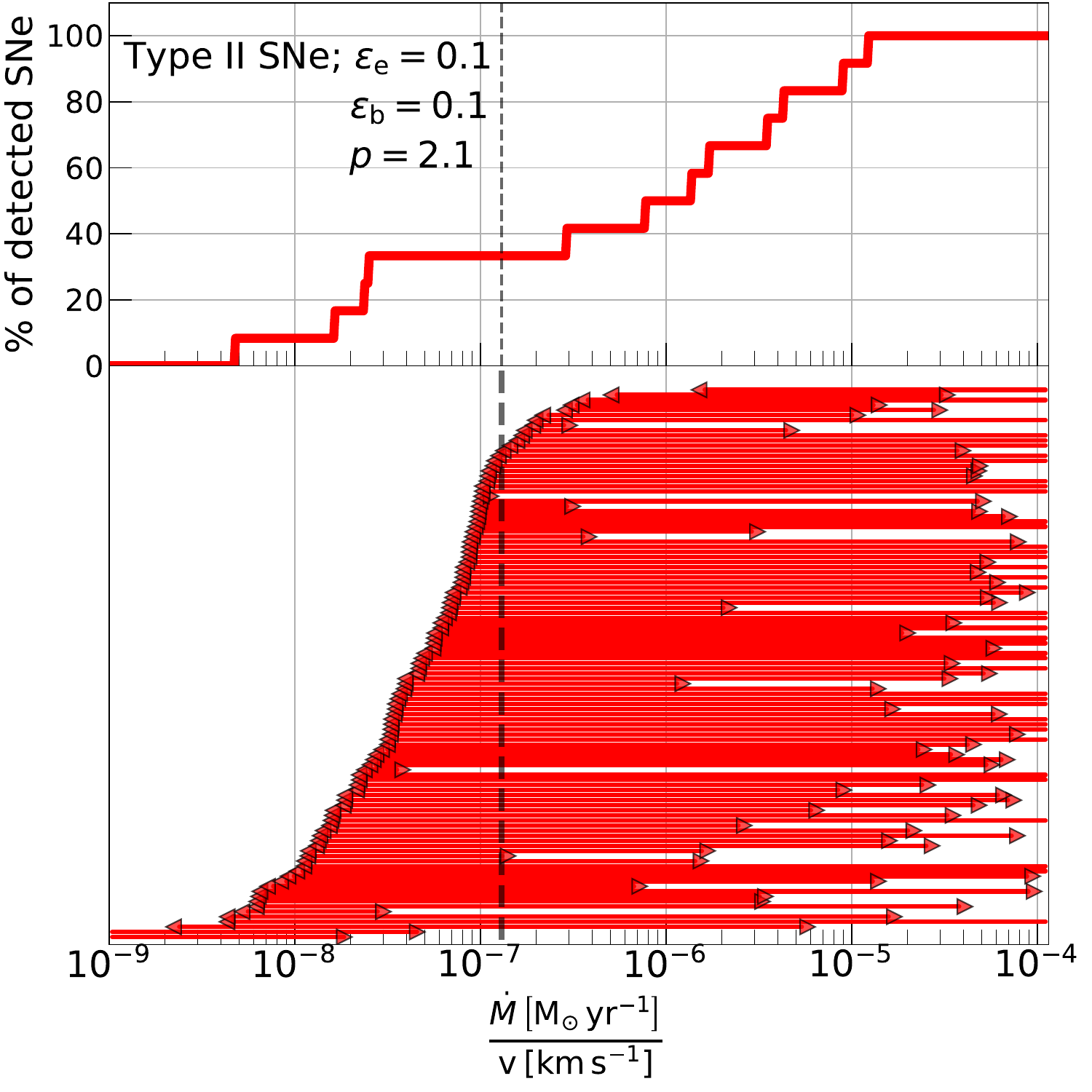"}
\includegraphics[width=0.43\linewidth]{"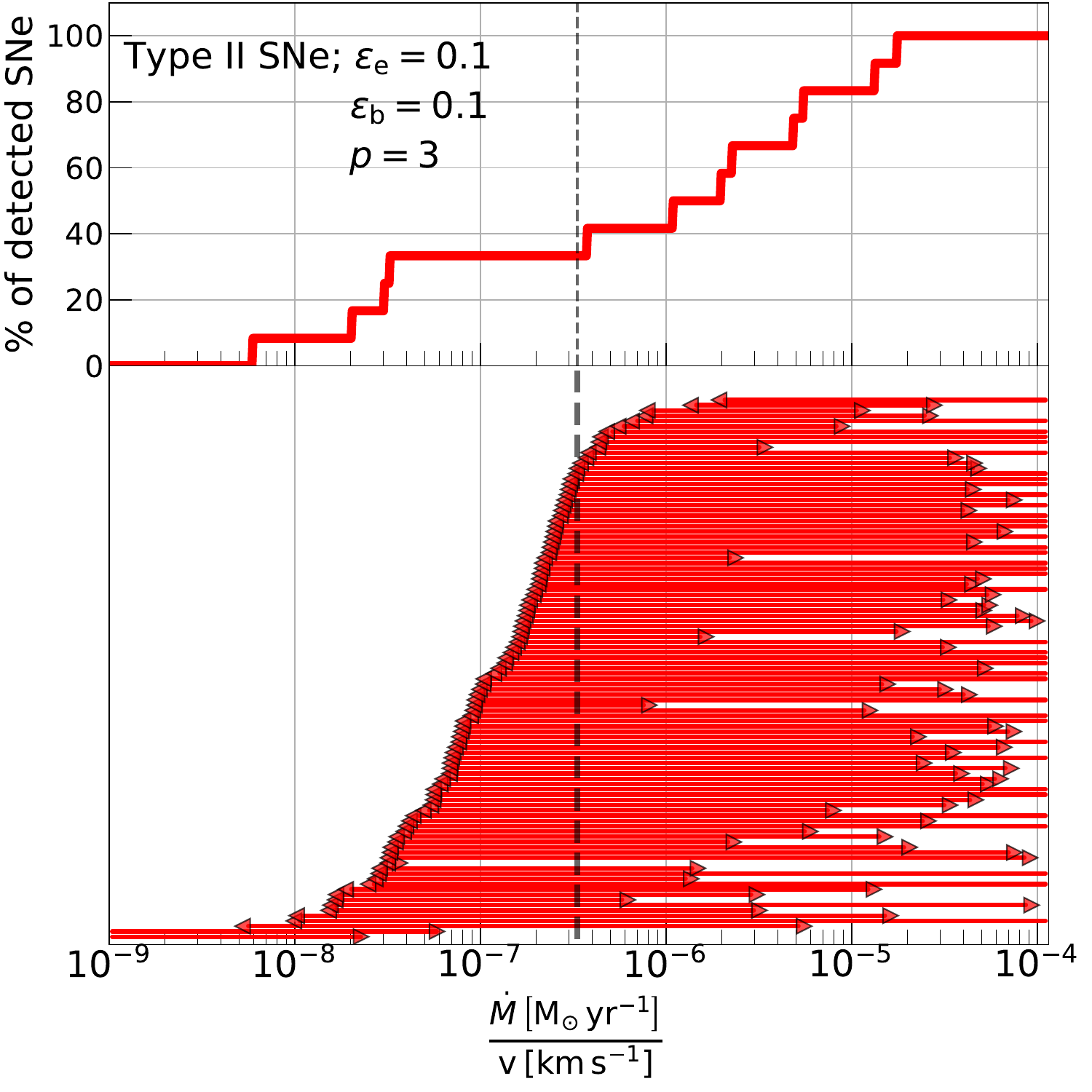"}
\caption{\footnotesize{Representation of the phase space of mass-loss rate for different assumptions on the micro-physical parameters. These plots are similar to the plots seen in Fig. \ref{fig: Forbidden regions combined} but the values of $p$ and $\epsilon_{\rm B}$ are varied to show the effect of changing our model assumptions. In the top left panel, we show the results with the same assumptions made in this paper. In the top right panel, we assume $p=2.4$ and $\epsilon_{\rm B} = 0.01$. In the two bottom panels, we assume $\epsilon_{\rm B} = 0.1$ and $p=2.1$ (left panel), and $p=3$ (right panel).}
\label{fig: Forbidden regions combined - model assumptions}}
\end{center}
\end{figure*}

In \S\ref{sec: caveats} we discussed the effects of changing the microphysical parameters of the shock on the inferred mass-loss rates. Below, we present the phase space of $\dot{M}/v_{\rm w}$ when assuming different values of the electron power-law index, $p$, and the fraction of energy that goes into the magnetic fields, $\epsilon_{\rm B}$ (see Fig. \ref{fig: Forbidden regions combined - model assumptions}). The total effect of reducing $\epsilon_{\rm B}$ by an order of magnitude (top right panel) compared to the original set of assumptions ($p=3$ for stripped-envelope SNe, $p=2.4$ for type II SNe, and $\epsilon_{\rm B}=0.1$) is shifting $\dot{M}/v_{\rm w}$ to higher values. This does not change our conclusion of the discrepancy between radio-non-detected SNe and SNe with an observed peak, however, the ruled-out region of $\dot{M}/v_{\rm w}$ does not rule low values suggested by theoretical models (as discussed in \S\ref{sec:summary} for the progenitors of type II-P and II-L SNe). As seen from the two bottom plots, setting $p=2.1$ (bottom left plot) shifts the entire phase space to lower values, i.e., we rule out lower values of $\dot{M}/v_{\rm w}$, while increasing $p$ to $3$ shifts the entire phase space to higher values of $\dot{M}/v_{\rm w}$. This implies that low values of the electron power-law index rule out even lower values of $\dot{M}/v_{\rm w}$, deepening the disagreement with mass-loss rates inferred from standard stellar evolution models \citep{Annu_rev_Smith_2014}. On the other hand, higher values of $p$ do not rule low values suggested by theoretical models (as was seen for $\epsilon_{\rm B}$ and discussed in \S\ref{sec:summary} for the progenitors of type II-P and II-L SNe).

\section{Data Tables}
\label{sec: tables}
\twocolumngrid

Table \ref{table:Limits_summary} provides a summary of the radio upper limits for non-detected SNe.
Table \ref{table:Peaks_summary} provides a summary of the peak flux densities of radio detected SNe. 
Each table is published in its entirety in the machine-readable format in the online Journal.
\begin{deluxetable}{cccc}[!hb]
\tablecaption{Radio upper-limits of CCSNe \label{table:Limits_summary}}
\tablehead{
\colhead{Name} &
\colhead{$\Delta t$} &
\colhead{$\nu$} &
\colhead{$F_{\rm \nu}$} \\
\colhead{} &
\colhead{[Days]} &
\colhead{[GHz]} &
\colhead{[mJy]} 
}
\startdata
SN1980O & 2589 & 4.86 & $<$0.36 \\ [0.5ex]
\hline
SN1982F & 1033 & 4.86 & $<$0.18 \\ [0.5ex]
 & 919 & 4.86 & $<$1.16 \\ [0.5ex]
\hline
SN1984E & 4121 & 1.425 & $<$0.073 \\ [0.5ex]
\hline
SN1985F & 488 & 4.86 & $<$0.33 \\ [0.5ex]
 & 7140 & 8.46 & $<$0.037 \\ [0.5ex]
 & 353 & 4.86 & $<$0.189 \\ [0.5ex]
\hline
SN1985G & 51 & 4.86 & $<$0.212 \\ [0.5ex]
 & 168 & 4.86 & $<$0.675 \\ [0.5ex]
 & 638 & 4.86 & $<$0.623 \\ [0.5ex]
\hline
SN1985H & 25 & 4.85 & $<$0.155 \\ [0.5ex]
 & 1025 & 4.86 & $<$0.3 \\ [0.5ex]
\hline
SN1987F & 1006 & 4.86 & $<$0.43 \\ [0.5ex]
 & 1363 & 4.8 & $<$0.18 \\ [0.5ex]
 & 1059 & 4.8 & $<$0.18 \\ [0.5ex]
 & 2132 & 4.9 & $<$0.09 \\ [0.5ex]
\hline
SN1987K & 2394 & 8.44 & $<$0.225 \\ [0.5ex]
 & 45 & 1.66 & $<$2.6 \\ [0.5ex]
 & 875 & 4.86 & $<$0.634 \\ [0.5ex]
\hline
SN1987M & 5891 & 8.46 & $<$0.034 \\ [0.5ex]
\enddata
\tablecomments{$\Delta t$ is the time in days since explosion, $\nu$ is the observed frequency in GHz, and $F_{\rm \nu}$ is the $3\sigma$ upper limit on the radio flux density.\\
Table \ref{table:Limits_summary} is published in its entirety in the machine-readable format. A portion is shown here for guidance regarding its form and content.
}
\end{deluxetable}

\begin{deluxetable}{cccc}
\tablecaption{Peaks of the radio emission \label{table:Peaks_summary}}
\tablehead{
\colhead{Name} &
\colhead{$\Delta t$} &
\colhead{$\nu$} &
\colhead{$F_{\rm p}$} \\
\colhead{} &
\colhead{[Days]} &
\colhead{[GHz]} &
\colhead{[mJy]} 
}
\startdata
SN1979C & 964.0 & 1.49 & 9.29 \\ [0.5ex]
\hline
SN1980K & 116.0 & 5.0 & 2.28 \\ [0.5ex]
\hline
SN1981K & 1480.0 & 1.5 & 1.28 \\ [0.5ex]
\hline
SN1983N & 13.3 & 5.0 & 36.7 \\ [0.5ex]
\hline
SN1984L & $<$112.0 & 1.5 & $>$1.01 \\ [0.5ex]
\hline
SN1985L & 335.0 & 4.86 & 0.897 \\ [0.5ex]
\hline
SN1986E & $<$245.0 & 4.86 & $>$0.304 \\ [0.5ex]
\hline
SN1986J & 1416.0 & 4.86 & 124.7 \\ [0.5ex]
\enddata
\tablecomments{When a peak in the light curve or spectrum is given, we estimate the peak flux density, time, and frequency as described in \S\ref{sec: detections}. When the peak is not observed we provide a lower limit on the peak flux density. If the lower limit is given from a spectrum we set the time of observation as the time of the peak, $\Delta t$, and provide a limit on the peak frequency, $\nu$. If the lower limit is given from a light curve we set the frequency of observation, $\nu$, as the peak frequency, and provide a limit on the time of the peak, $\Delta t$.\\
Table \ref{table:Peaks_summary} is published in its entirety in the machine-readable format. A portion is shown here for guidance regarding its form and content.
}
\end{deluxetable}

\clearpage
\bibliography{Bib}

\end{document}